%% file: CLICdp_eventshapes.tex
\documentclass[11pt,a4paper]{scrartcl}
\usepackage{CLICdp}

\usepackage{CLICdp_definitions}

\usepackage{hyperref}

\hypersetup{
    bookmarksnumbered=true,   % put section numbers in bookmarks
    breaklinks=true,          % line breaks in links
    pdftoolbar=true,          % show Acrobat toolbar?
    pdfmenubar=true,          % show Acrobat menu?
    pdffitwindow=false,       % window fit to page when opened
    pdfstartview={FitV},      % fits the height of the page to the window
    pdfpagelayout=SinglePage, % show only one page in pdf viewer
    pdfnewwindow=true,        % links in new window
    colorlinks=true,          % false: boxed links; true: colored links
    linkcolor=darkgreen,      % color of internal links (change box color with linkbordercolor)
    citecolor=darkred,        % color of links to bibliography
    filecolor=darkmagenta,    % color of file links
    urlcolor=darkblue         % color of external links
}

\usepackage{epstopdf}

\title{Effect of PYTHIA8 tunes on event shapes and top-quark reconstruction in $e^+e^-$ annihilation at CLIC
}

\clicdpnote{2017}{005}

\date{\today}
\addauthor{S.~Chekanov\thanks{chekanov@anl.gov}}{\institute{1}}
\addauthor{M.~Demarteau}{\institute{1}}
\addauthor{A.~Fischer}{\institute{1}\institute{2}}
\addauthor{J.~Zhang}{\institute{1}}

\addinstitute{1}{HEP Division, Argonne National Laboratory,
9700 S.~Cass Avenue, Argonne, IL60439, USA 
}

\addinstitute{2}{
College of Arts and Sciences, Loyola University Chicago, 1032 W. Sheridan Road, Chicago IL60660, USA
}

\abstract{
This paper describes the effect of PYTHIA8 tunes on event
simulation of $e^+e^-$ collisions with center-of-mass (CM) energies
of 380 GeV and 3 TeV
at the proposed CLIC collider.
Event shapes, such as thrust, thrust major, thrust minor, oblateness, as well as particle multiplicities
have been analyzed and relative differences with respect to the default PYTHIA8 tune were determined.
The effect of tunes on top-mass reconstruction in the resolved and boosted regimes was analyzed.
No statistically significant variation  for reconstructed top masses using invariant masses of
three jets  was found for events with a CM energy of 380~GeV.
For the fully boosted top reconstruction at a CM energy of 3~TeV,
a  significant shift in reconstructed top mass of about 700~MeV for the ``Montull'' tune was observed.
This shift correlates with an increase in particle multiplicity compared to all
other PYTHIA8 tunes.
}

\titlecomment{This work was carried out in the framework of the CLICdp Collaboration} %use either this or /onbehalfof above

\graphicspath{ {./logos/}{./figs/} }

\begin{document}

% generates the title page
\titlepage

% include source for sections
\input{./maintext.tex}

\input{./eventshapes.tex}

\input{./topreconstruction.tex}

\input{./conclusion.tex}

\section*{Acknowledgements}
We thank P.~Skands and T.~Sjostrand for discussions.
This research was performed using resources provided by the Open Science Grid,
which is supported by the National Science Foundation and the U.S. Department of Energy's Office of Science.
We gratefully acknowledge the computing resources provided on Blues,
a high-performance computing cluster operated by the Laboratory Computing Resource Center at Argonne National Laboratory.
Argonne National Laboratory's work was supported by the U.S. Department of Energy, Office of Science under contract DE-AC02-06CH11357.

\newpage
% add references
\printbibliography[title=References]

\newpage
\input{./papfigures.tex}

\end{document}

%% file: maintext.tex
%\newcommand{\latex}{\LaTeX\xspace}
%\lstset{defaultdialect=[LaTeX]TeX}

%%%%%%%%%%%%%%%%%%%%%%%%%%%%%%%%%%%%%%%%%%%%%%%%%%%%%%%%%%%%%%%%%%
\section{Introduction}
%%%%%%%%%%%%%%%%%%%%%%%%%%%%%%%%%%%%%%%%%%%%%%%%%%%%%%%%%%%%%%%%%%

High-precision measurements at future $e^+e^-$ collisions require a good understanding 
of hadronisation, which involves non-perturbative QCD physics.
Currently, modeling of this stage in particle production can be successfully achieved using Monte Carlo simulations.
Such simulations have a number of free parameters that have to be tuned
to experimental data using different techniques. This can lead to uncertainties that can affect final measurements when such 
Monte Carlo simulations are used for unfolding detector effects and various corrections.

The effect of Monte Carlo tunes on hadronic final states in $e^+e^-$ annihilation
at the CLIC collider \cite{clic} can be studied using the PYTHIA8 Monte Carlo generator \cite{Sjostrand:2007gs}. 
It has seven tunes dealing with hadronization, particle (flavor) composition and time-like showering aspects.
The tunes use different sets of the LEP1-LEP2 data.
The default setting used by PYTHIA8 is the so-called Monash 2013 \cite{Skands:2014pea} tune,
which can be applied to both $e^+e^-$ and $pp$/$p\bar{p}$ data.
More information on other PYTHIA8 tunes can be found in Ref.\cite{Sjostrand:2007gs}.

This paper discusses the PYTHIA8 tunes and their impact on hadronic-final states 
in $e^+e^-$ annihilation at the 
centre-of-mass (CM) energies  planned for the CLIC collider.
In particular, effects on top-mass reconstruction were studied using the 
resolved and boosted top reconstruction
techniques.

\section{Simulations}
In this study two million events per PYTHIA8 
tune were generated and compared against the Monash 2013 tune (default).
Events were simulated for two CM energies, 380~GeV and 3~TeV.
For the event-shape studies, 
the generated 
processes are  $e^+e^-\rightarrow Z^*/\gamma \rightarrow q\bar{q}$, where $q$ ($\bar{q}$) are quark (anti-quark), which
includes all quark flavors but the top quarks.
To study top reconstruction, the events were generated for the process 
$e^+e^-\rightarrow Z^*/\gamma \rightarrow t\bar{t}$,
where one top quark decays to $W\> b$, followed by  $W^{\pm} \rightarrow l^{\pm}\nu$,
while the second $W$ decays to two jets.
The initial-state photon radiation was turned off for the event shape studies, but switched on for top reconstruction studies
to obtain the most realistic case for top-quark reconstruction.
Event shapes and jets were reconstructed from stable particles, which are
defined as having a lifetime  $\tau$  larger than $3\cdot 10^{-10}$ seconds. Neutrinos were excluded from consideration.
No kinematic cuts have been applied to stable particles. 

In this paper, tunes for different simulations 
will be indicated as "$T[N]$", where "$[N]$" is an integer number corresponding to the PYTHIA8 parameter  
``Tune:ee'' which specifies the implemented tune used to generate events. 
In the following, the default (``Monash 2013'') tune is indicated with $T0$.
Other tunes, with  $N$ running from 1 to 6, are described in \cite{Sjostrand:2007gs}.

All files generated for this study are available from the HepSim public repository \cite{Chekanov:2014fga}.
The CLIC detector response was not simulated, thus the effect of the tunes on reconstructed variables gives an upper
limit of the possible impact on event shapes and top-mass measurements, 
without expected smearing due to detector effects and beam backgrounds.

%% file: eventshapes.tex
\section{Effect from tunes on hadronic final states}
\label{sec:eventshapes}

\subsection{Multiplicity distributions}

One simple variable that is potentially sensitive to hadronisation and particle-event composition
is particle multiplicity per collision event.
The study of the effect of PYTHIA8 tunes on particle multiplicity was 
done using the $e^+e^-\rightarrow Z^*/\gamma \rightarrow \mathrm{hadrons}$
process, excluding top production. 
Figure~\ref{fig:multiplicity1} shows the neutral- and charged-particle multiplicity distributions 
for 380~GeV and 3~TeV CM energy, respectively.
The result indicates  that the tune $T2$, which corresponds
to the ``Montull'' setting, shows the largest increase in particle multiplicity. 
A similar observation was made using the multiplicity distribution of charged particles (not shown).

Figure~\ref{fig:multiplicity2} shows the distribution of particle multiplicity for  380~GeV and 3~TeV CM  
energy for the  $e^+e^-\rightarrow Z^*/\gamma \rightarrow t\bar{t}\mathrm{\>(all\> decays)}$ process.
The conclusion stays the same as in the case of inclusive production, i.e. the effect
of tunes can be as large as 100\% for the large multiplicity tail, with the largest
enhancement coming from the Montull tune. 

\subsection{Event shapes}

Event shape variables are used for studies of  global features of particle flows in events.
Studies of the energy flows in hadronic final states of $e^+e^-$ annihilation
have allowed tests of perturbative QCD. 
Some of the event shape variables have been discussed in the recent review \cite{eventshapes}. This study uses 
several popular variables defined below:  

\begin{itemize}
 
\item  Thrust of an event is defined as $T = \max_{\vec{n}} \frac{\sum_{i} |\vec{p}_{i}\cdot\vec{n}|}{\sum_{i}|\vec{p}_{i}|}$, where
$|\vec{n}| = 1$. The sum runs over three-momenta of all final state particles (charged and neutral). The thrust axis 
is defined by the vector $\vec{n}_T$ for which the maximum is obtained. 

\item Thrust major is defined as $T_M = \max_{\vec{n}_M} \frac{\sum_{i} |\vec{p}_{i}\cdot\vec{n}_M|}{\sum_{i}|\vec{p}_{i}|}$,
assuming the additional condition that  $\vec{n}_{M}$ should lie in the plane perpendicular to $\vec{n}_T$; 

\item Thrust minor is defined as
$T_m = \frac{\sum_{i} |\vec{p}_{i}\cdot\vec{n}_m|}{\sum_{i}|\vec{p}_i|}$, where  $\vec{n}_m = \vec{n}_T \times \vec{n}_M$.
The thrust minor is perpendicular to both the thrust and the major axis.

\item Oblateness, $O_{bl} =T_M-T_m$. 

\end{itemize}

For convenience, we use the $1-T$ variable, instead of $T$.  Such a  re-definition leads to a falling 
distribution, similar to the other three event-shape variables, $T_M$, $T_m$ and $O_{bl}$. 
In the limit of two narrow back-to-back jets, $1-T$ is close to 0, 
while a value of 1/2 corresponds to events with a uniform distribution of momentum flow in all directions.

Figure~\ref{fig:thrust} shows the distribution of the trust values $1-T$ for two different CM energies of the
 $e^+e^-\rightarrow Z^*/\gamma \rightarrow \mathrm{hadrons}$  process.
It indicates that the $T2$ (``Montull'') setting 
shows the largest  deviation from the central tune in the limit of small values of $1-T$.
This deviation is close to $4-5\%$ for the ratio with default tune. 
Large values of the  $1-T$ distribution suffer from low statistics, therefore, it is difficult to make
a conclusive statement on contributions from different tunes. 

Figure~\ref{fig:major} shows the values of the major thrust for 380~GeV and 3~TeV CM energies.
Similarly, Fig.~\ref{fig:minor} shows the values of the minor thrust. 
The difference between different tunes was found to be smaller compared to the 
multiplicity distributions discussed before, with the largest
deviation of $10\%$ for the $T2$ tune.

Figure~\ref{fig:oblateness} shows the distribution of the values of oblateness for different CM energies.
As in the case of major and minor thrust values, no significant effect from the use of different tunes was found. 

Figures~\ref{fig:thrust_ttbar}, \ref{fig:major_ttbar}, \ref{fig:minor_ttbar} and \ref{fig:oblateness_ttbar} show
the event-shape variables for the  $e^+e^-\rightarrow Z^*/\gamma \rightarrow t\bar{t}\mathrm{(all\> decays)}$ process.
As expected, these distributions 
for top production have different shapes compared to 
the $e^+e^-\rightarrow Z^*/\gamma \rightarrow \mathrm{hadrons}$  process,
indicating  more uniform distributions of momentum flow in all directions.
As before, the effect of different tunes was found to be smaller compared to the
multiplicity distributions shown in Fig.~\ref{fig:multiplicity2}.
The observed variations in the event shape distributions 
are at the level of $5-10\%$ compared to the default tune.

%% file: topreconstruction.tex
\section{Effect on top reconstruction}
\label{sec:top}

\subsection{Resolved reconstruction}

The effect of the PYTHIA8 settings described in the previous section was studied in the context 
of top reconstruction.  
Semi-leptonic $t\bar{t}$ decays in $e^+e^-$ annihilation were selected as candidates for three-jet mass 
reconstruction using events with a lepton from the $W$ decays. 
The jets were
reconstructed using the anti-$k_T$ algorithm \cite{Cacciari:2008gp} as implemented in the FastJet package~\cite{fastjet}. 
The input for the jet reconstruction excludes neutrinos  and the lepton from the $W$ decays.
This jet algorithm used a distance parameter of $R=1.0$. 
Note that the recent CLIC studies \cite{Seidel:2013sqa} of the top reconstruction in $e^+e^-$ at 500~GeV suggest even a larger jet size ($R=1.3$). 
For the generator-level studies presented in this paper, we did not find significant improvements in top reconstruction for larger jet sizes. 

Hadronic jets were reconstructed  in the exclusive mode by requiring exactly four jets in an event.
Jets that originate from $b-$quarks were identified by matching jet four-momenta with
the momenta of $b-$quarks from the truth-level particle record using a cone algorithm defined in rapidity ($y$) and the azimuthal angle ($\phi$)
with a size of $0.3$.  
When reconstructing the top mass for the 380~GeV events, light flavoured jets were merged to build the $W$-boson mass. 
The two-jet mass combinations 
were then constrained to fall within 20~GeV of the nominal $W$ mass value of 80.385 GeV. 
Such two-jet combinations correspond to $W$ boson candidates.
b-jets that shared a mother particle with the leptonic 
$W$ were excluded,  and the remaining $b$-jets were combined with the $W$-boson candidates 
to calculate the top mass.

Figure~\ref{fig:resolv}(a) shows the dijet invariant mass distribution for light-flavour jets.
A clear peak at around 80~GeV was observed, which corresponds to the $W$ boson mass.
The effect of different tunes is indicated in the ratio plot to the default PYTHIA8 tune.

Figure~\ref{fig:resolv}(b) shows the reconstructed mass of three-jet combinations, 
after combining the $W$-boson candidates  
reconstructed from two jets with the third jet identified 
as the $b-$jet.
This figure shows that the effect from different tunes on this distribution is negligible.
For a quantitative assessment of the mass shifts, a Breit-Wigner function was used to fit the three-jet mass distribution around the
peak position of this distribution. 
Generally, the fit function does not describe well the asymmetric distribution shown in Figure~\ref{fig:resolv}(b). 
Nevertheless,
restricting the fit range to $\pm 4$~GeV around the peak leads 
to a good value of $\chi^2/\mathrm{ndf}\simeq 1$,  
and helps a quantitative assessment of shifts in the peak position. 
It was determined that the shifts in the  peak position of the  Breit-Wigner distribution 
with respect to the default PYTHIA8 tune are within 80~MeV for all implemented tunes.  
At this moment, this value has a non-negligible contribution from statistical uncertainties, therefore,
the actual effect can be smaller than this value.

\subsection{Boosted reconstruction}

The resolved top reconstruction described above has a low reconstruction efficiency for 
the boosted regime at 3~TeV CM energy. 
For this large CM energy,  the transverse momenta of top quarks are 
large enough that $b$-jets overlap with light-flavour jets making the differentiation of light-flavour and $b-$jets difficult.

For the boosted-top studies, jets were
reconstructed using the anti-$k_T$ algorithm \cite{Cacciari:2008gp}
with  a distance parameter of $R=1.2$.
The events were forced to have two jets.
A larger jet size of $R=1.2$, compared to $R=1.0$ for the resolved case, 
was found to have a better reconstruction efficiency in the boosted regime.
This effect is expected since the larger jet size is more appropriate for collecting the decay products of boosted top quarks. 
Then, the mass of the leading jet was calculated for each PYTHIA8 tune.

Figure~\ref{fig:threejet} shows the mass of the leading jets in the boosted regime. As for the resolved case, the peak position
for each PYTHIA8 tune was determined using the fits with the Breit-Wigner distribution.
Generally, the Breit-Wigner distribution is too simple for the description  
of the asymmetric jet-mass distribution.
However, it is sufficient for quantitative estimates of  the observed shift after restricting the
fit range to $\pm 4$~GeV around the peak position of the jet mass distribution.   
It was found that the largest shift for jet mass with respect to the PYTHIA8 default is due to the Montull tune.
This shift is about 700~MeV.
The observed large effect on jet masses is not totally surprising since our previous studies
have indicated the significant effect of the Montull tune on particle multiplicities, and 
jet masses are known to be sensitive to the number of jet constituents.

%% file: conclusion.tex
\section{Conclusions}

We have performed an analysis of PYTHIA8 tunes for $e^+e^-$ annihilation and their effect on several event
shape distributions and on top-mass reconstruction using resolved and boosted techniques.
This study was performed for two CM energies, 380~GeV and 3~TeV,  anticipated for the CLIC collider.

Most tunes of PYTHIA8 show small effects on hadronic final states of $e^+e^-$.
The largest change on the reconstructed variables
was found for the Montull tune to the LEP1 particle composition.
The effect is significant  for
multiplicity distributions,
where the difference between the Montull and the default (``Monash'') parameter sets
can be as large as 100\% for events with large
multiplicities of final-state particles.
The Montull tune  is  directed towards the flavour composition, and  care should be taken considering it
for future studies since it is rather outdated \cite{torbjorn}. 

Compared to the multiplicity distributions, the effect of PYTHIA8 tunes is significantly smaller for the event shape variables considered
in this paper.
A typical effect of about $5\%$ was found for small values of $1-T$.
The Montull tune shows the largest shift (of about 10\%) for the  minor and major thrust values.
However, it is difficult to determine systematic trends for large event-shape values due to limited  statistics.

The reconstruction of the top mass from dijets in semi-leptonic top decays
using different PYTHIA8 tunes does not indicate significant mass shifts with respect to the
default PYTHIA8 settings.  The observed mass shifts were found to be less than $80$~MeV
at the CM energy of 380~GeV.

For the boosted top reconstruction using 3~TeV CM energy, a shift in jet mass
from boosted top decays was found to be of the order of 700~MeV. This shift originates from the use of the Montull  tune,
and it is qualitatively consistent with the observed  increase in particle multiplicity for this tune.
If the Montull tune is used for future CLIC studies, the observed mass shift can have an impact on various searches
and other measurements in the boosted regime.

%% file: papfigures.tex
\section{Figures}
\label{sec:figures}

\begin{figure}[ht!] 
\centering
  \includegraphics[width=0.6\textwidth]{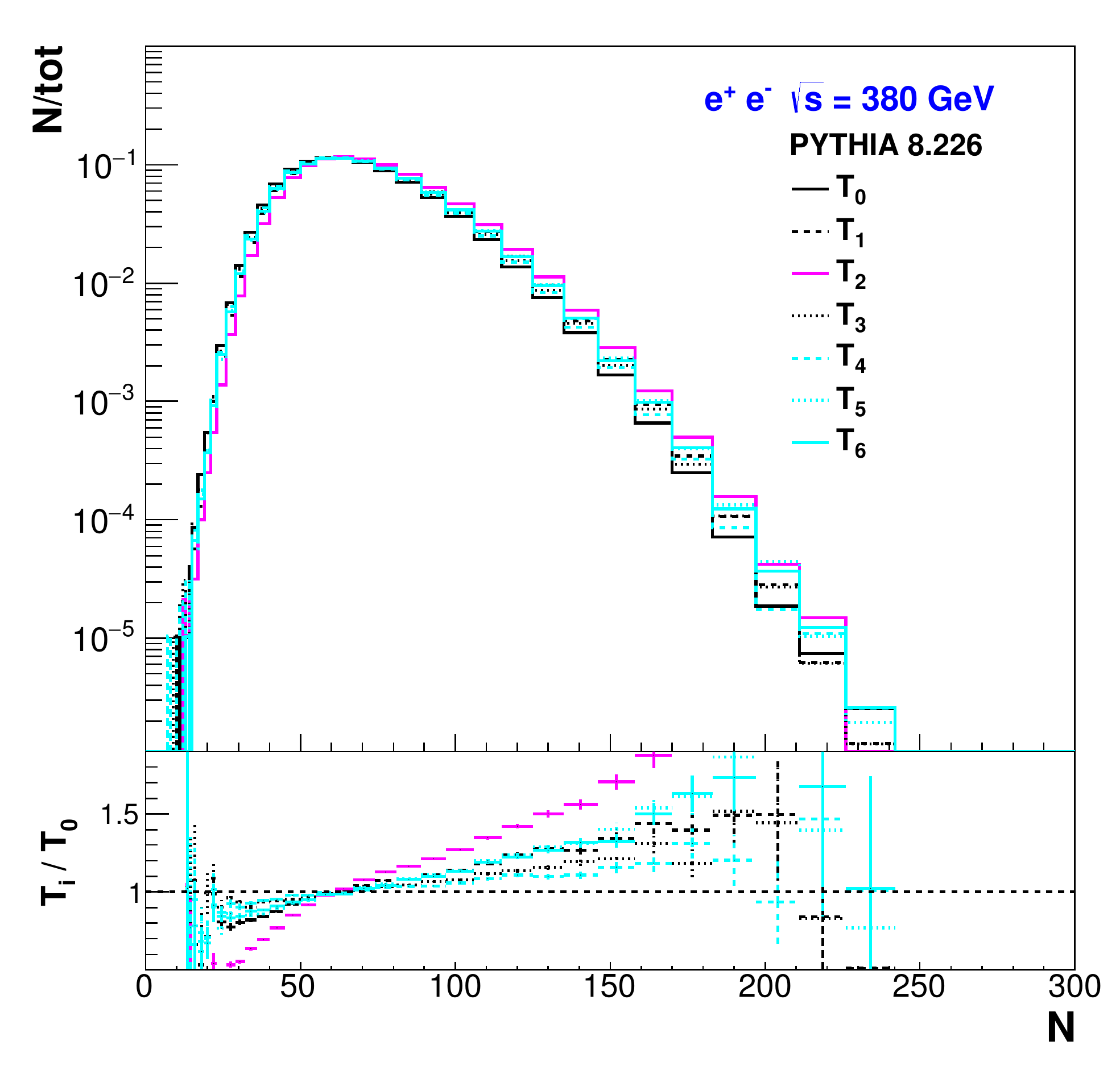}
  \includegraphics[width=0.6\textwidth]{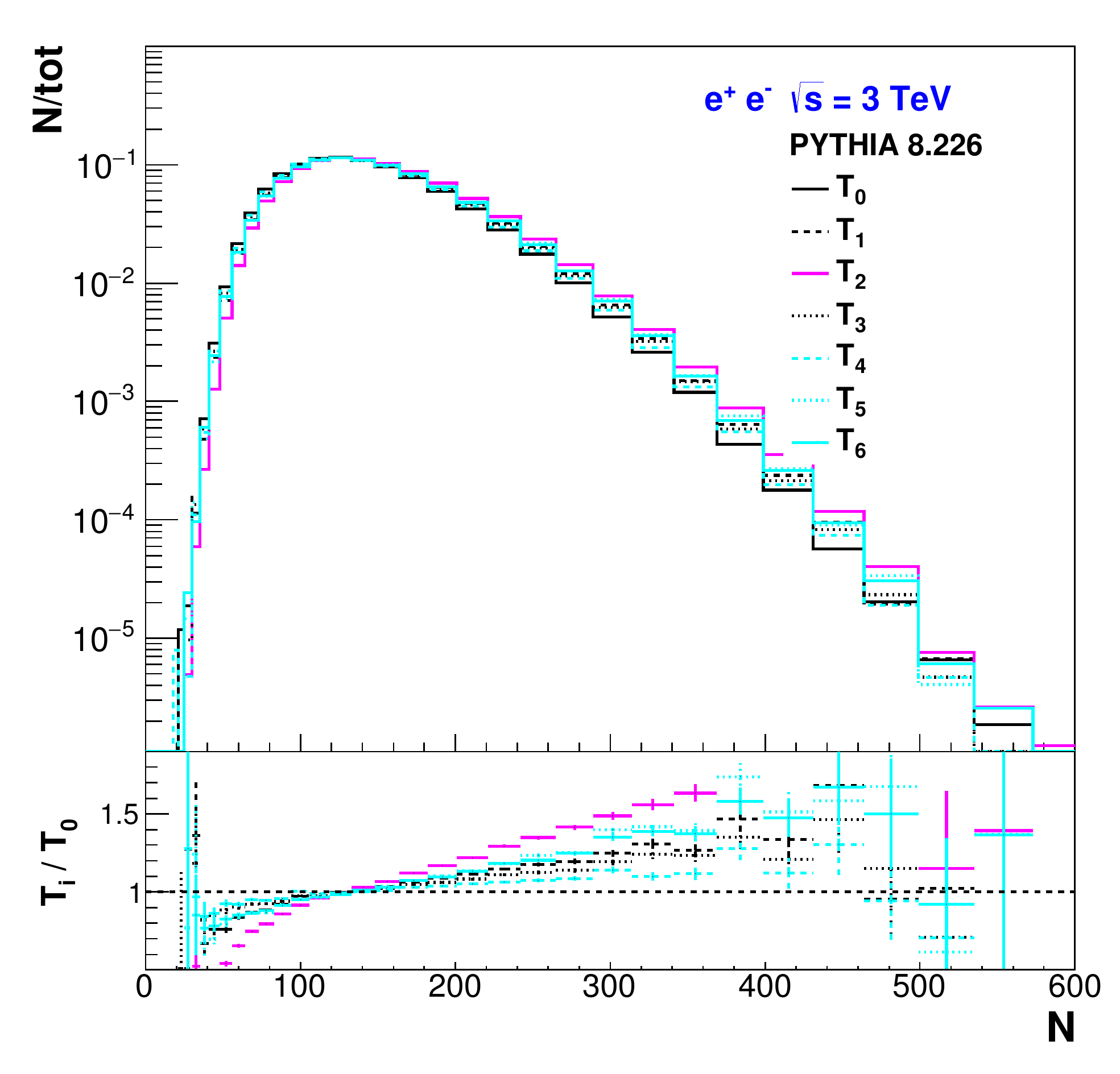}
\caption{The distribution of particle multiplicity (including charged and neutral particles)  
for different PYTHIA8  tunes at 380 GeV and 3~TeV CM energies of $e^+e^-$ annihilation with hadronic decays of $Z^*/\gamma$ (excluding $t\bar{t}$). The bottom plot
shows the ratios of different tunes with the default $T0$ (``Monash'') tune. }
\label{fig:multiplicity1}
\end{figure}

\begin{figure}
\centering
  \includegraphics[width=0.6\textwidth]{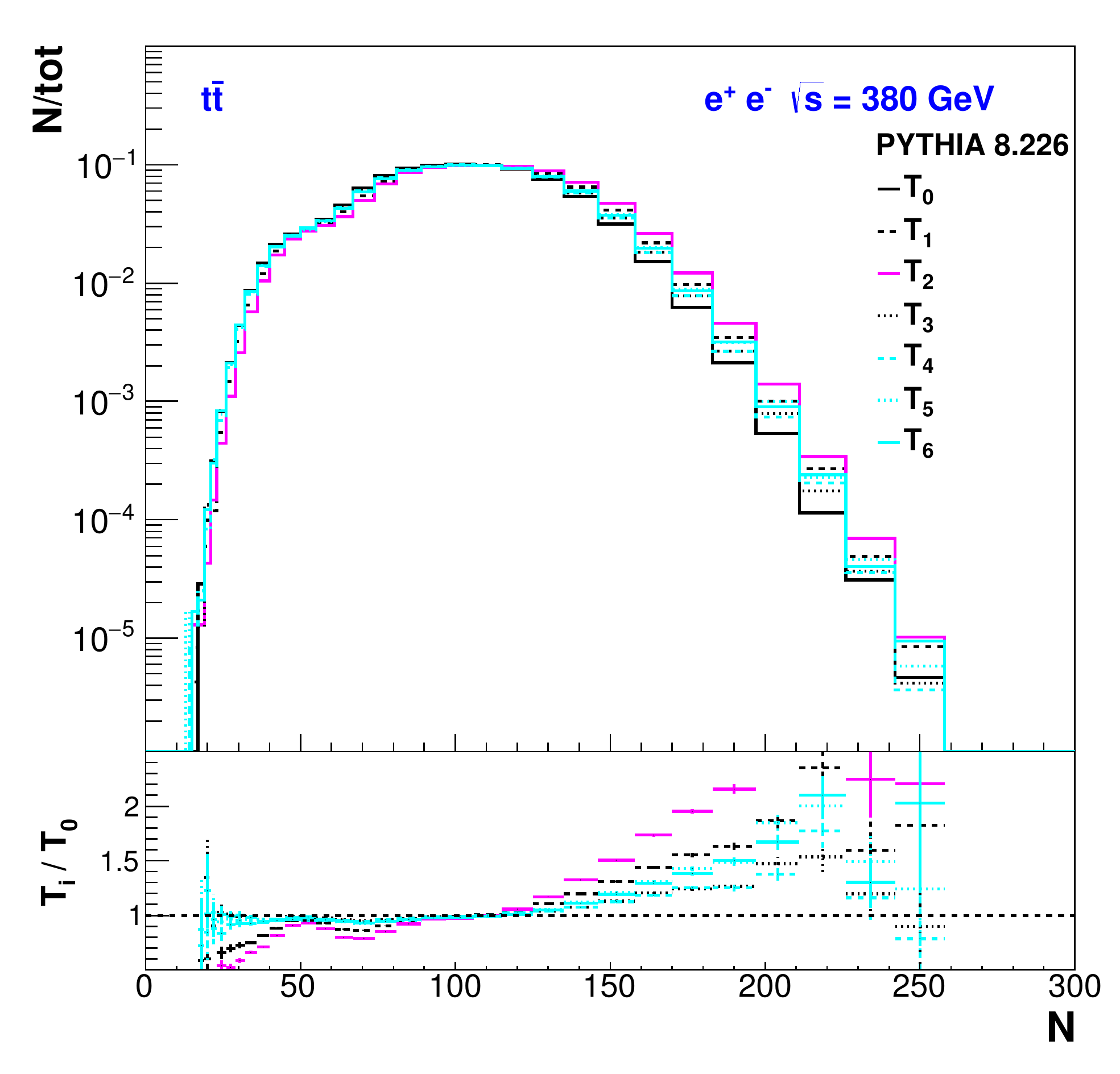}
  \includegraphics[width=0.6\textwidth]{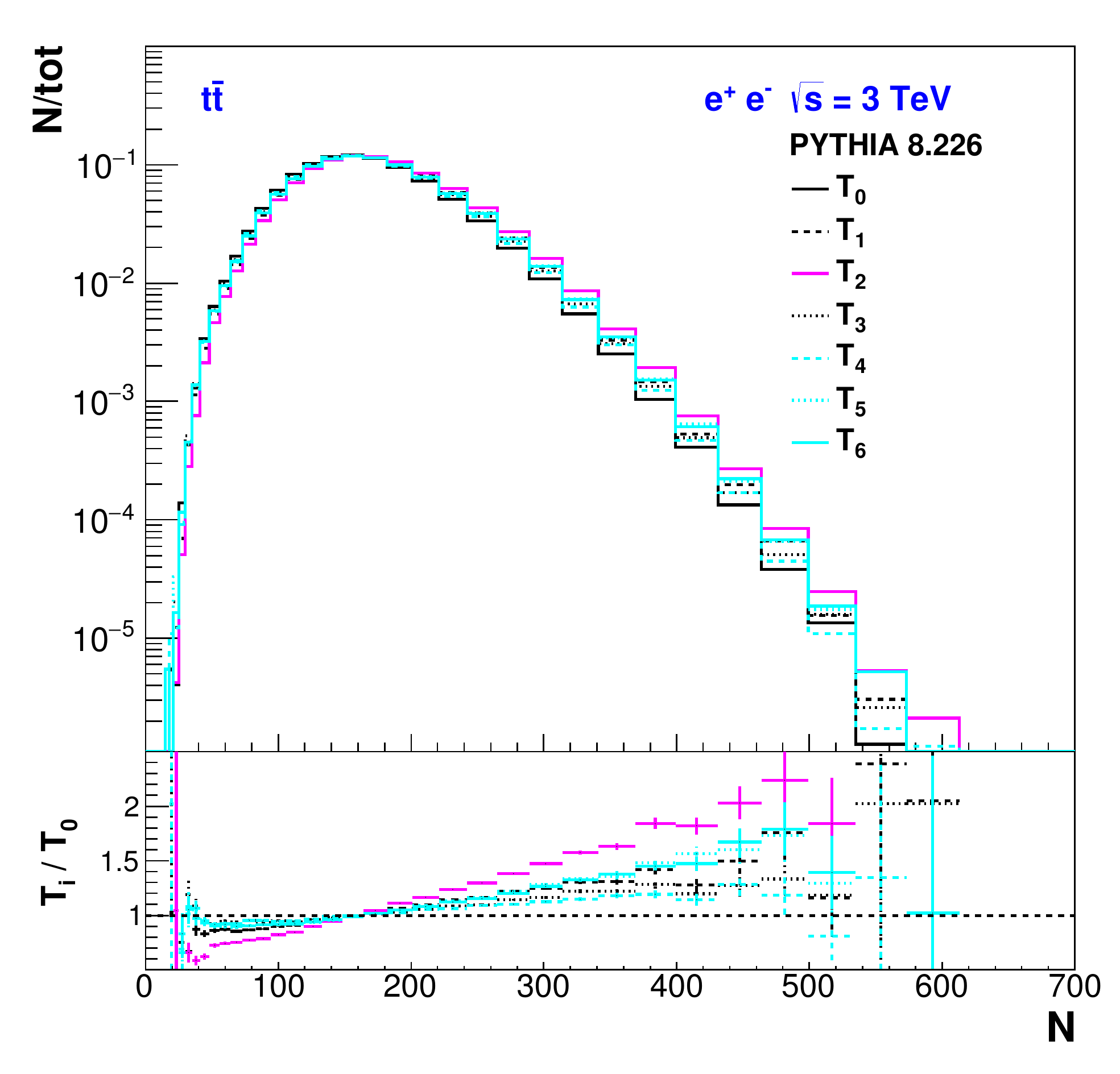}
\caption{The distribution of particle multiplicity (charged and neutral) for  different PYTHIA8  tunes at 380 GeV and 3~TeV CM energies 
in $e^+e^-\rightarrow Z^*/\gamma \rightarrow t\bar{t}\mathrm{\>(all\> decays)}$.}
\label{fig:multiplicity2}
\end{figure}

%%% inclusive QCD
\begin{figure}
\centering
  \includegraphics[width=0.6\textwidth]{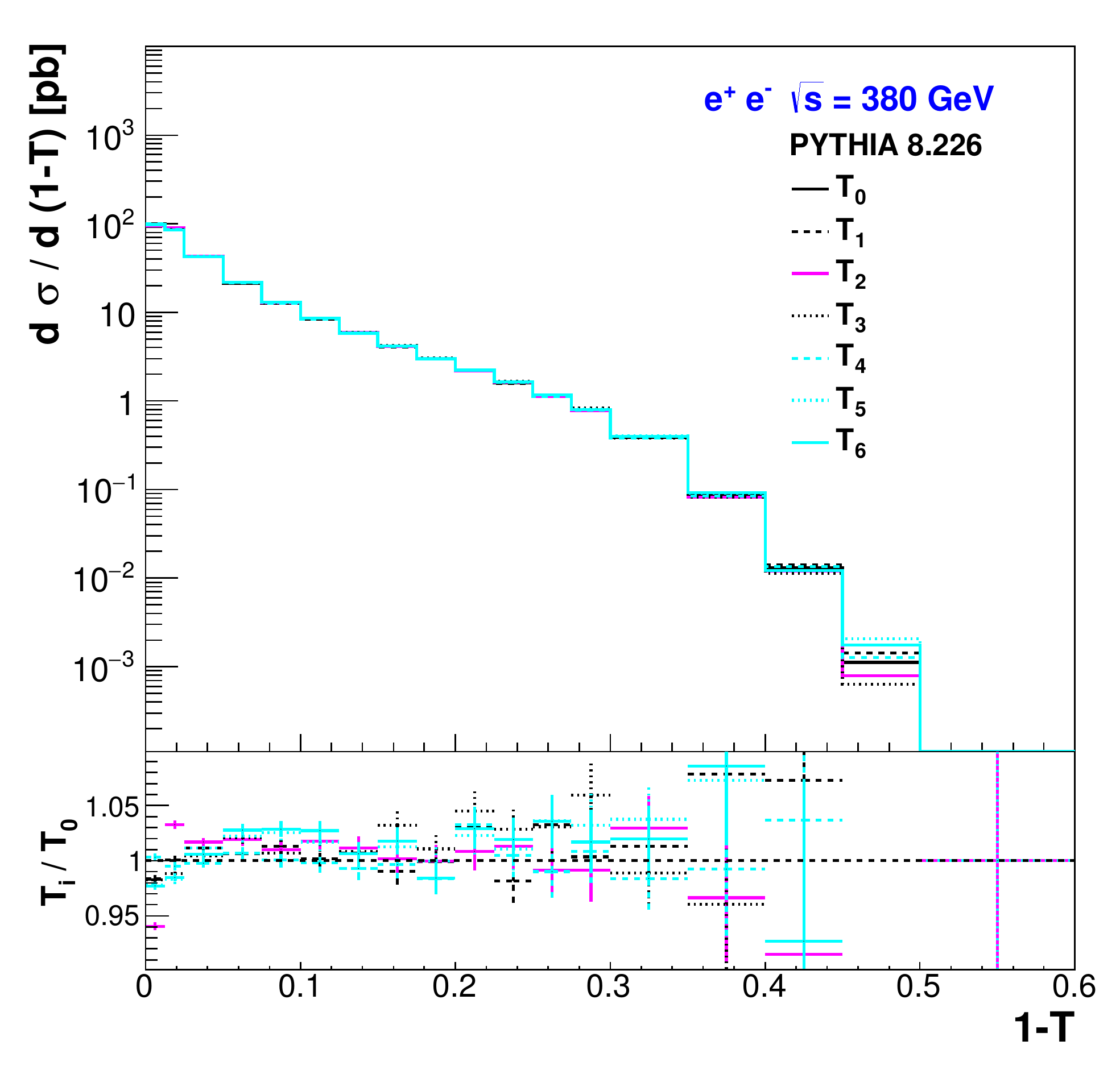}
  \includegraphics[width=0.6\textwidth]{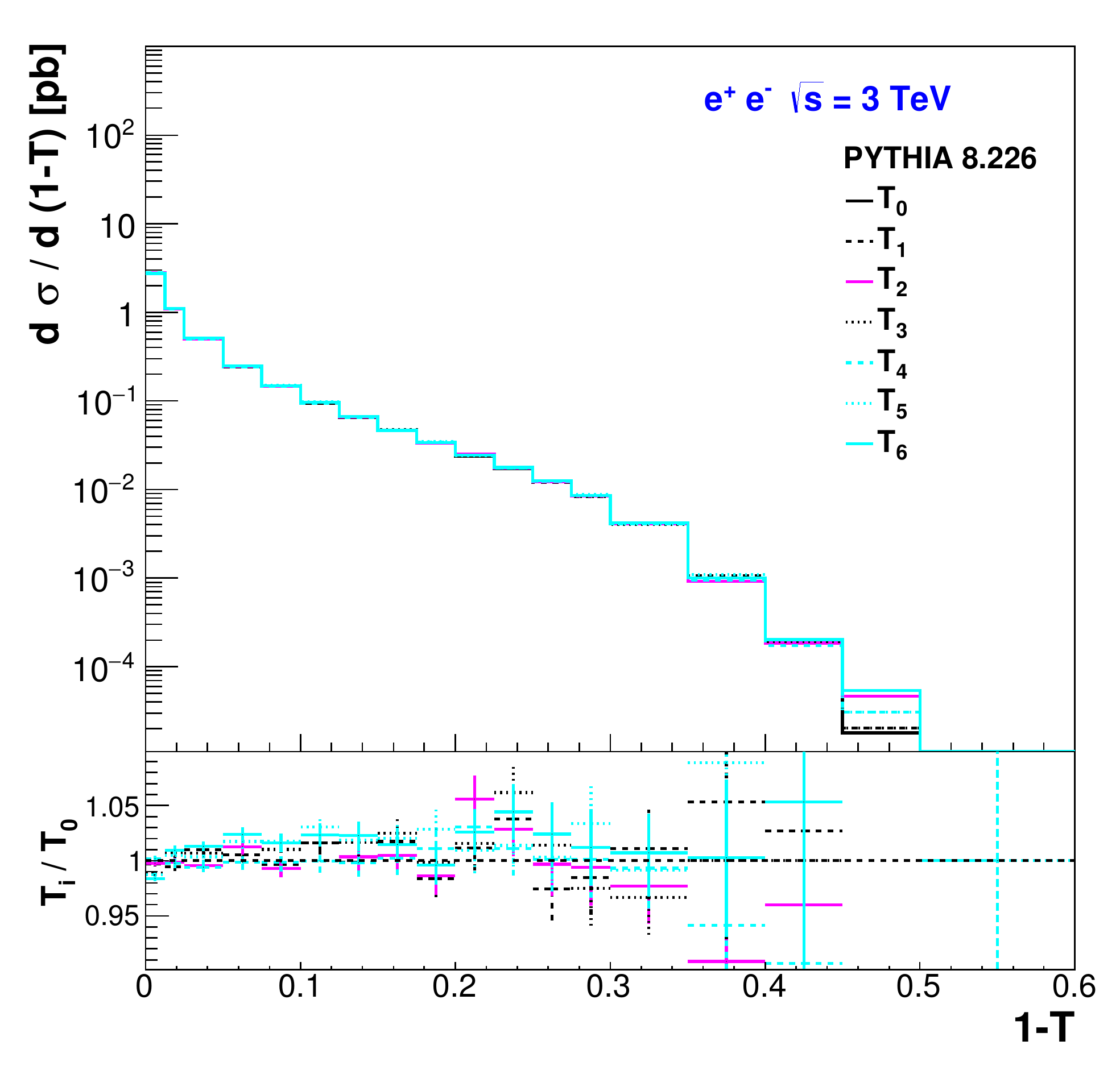}
\caption{The distribution of the thrust values ($1-T$) for different PYTHIA8 tunes at 380 GeV and 3~TeV CM energies in 
$e^+e^-\rightarrow Z^*/\gamma \rightarrow \mathrm{hadrons}$ (excluding $t\bar{t}$).}
\label{fig:thrust}
\end{figure}

\begin{figure}[!ht]
\centering
  \includegraphics[width=0.6\textwidth]{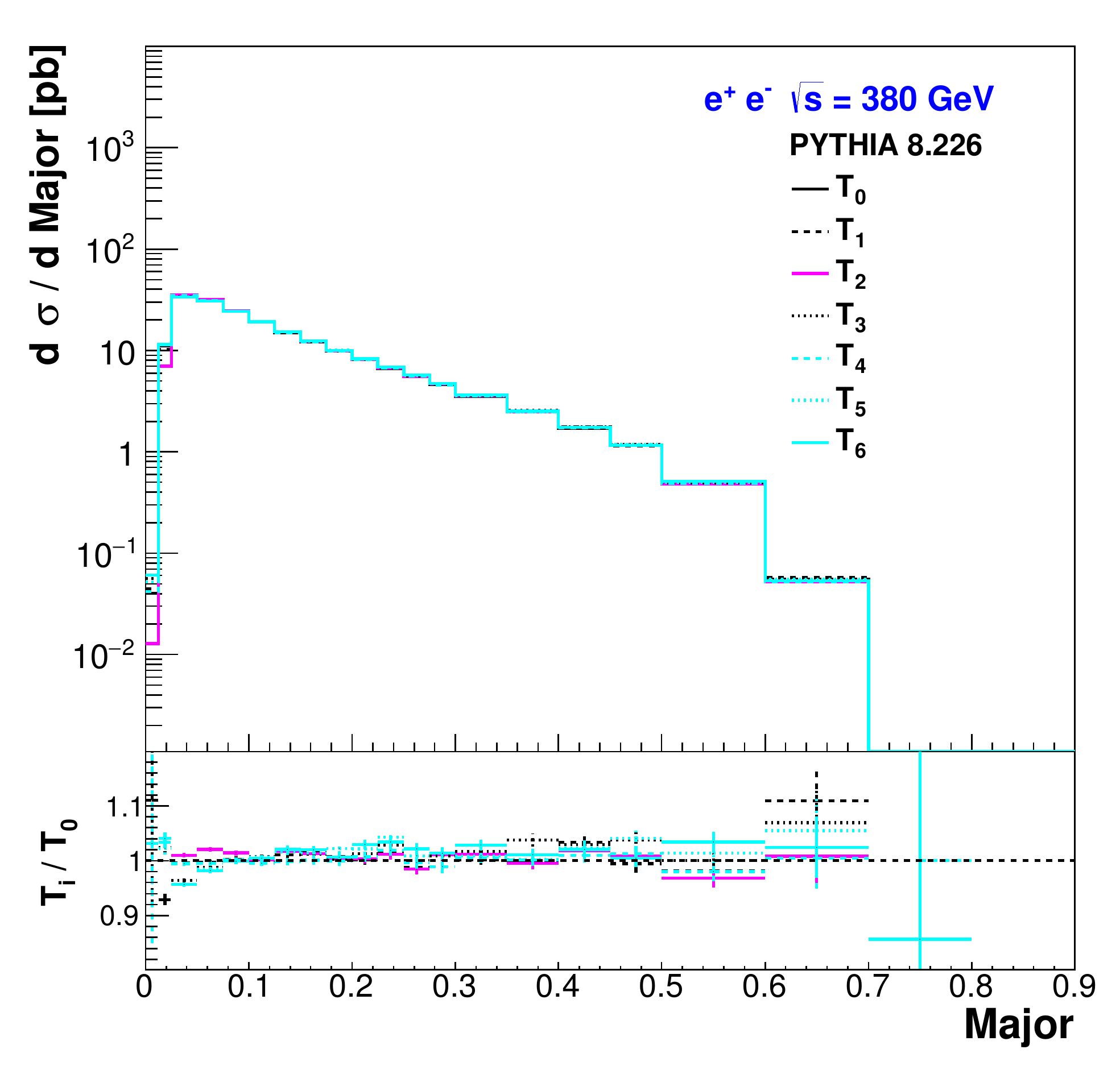}
  \includegraphics[width=0.6\textwidth]{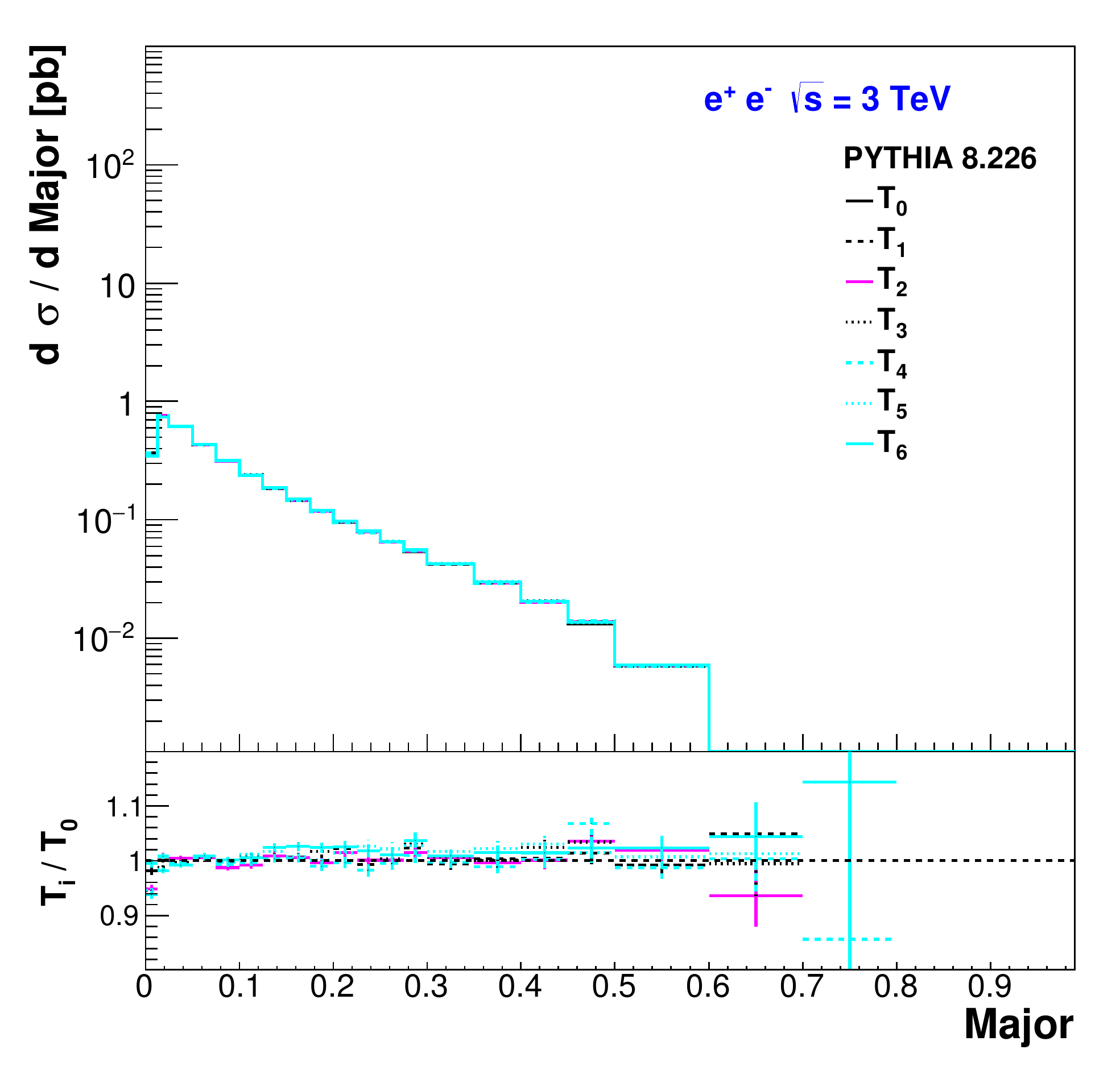}
\caption{The distribution of the major thrust values for different PYTHIA8 tunes at 380 GeV and 3~TeV CM energies  in  $e^+e^-\rightarrow Z^*/\gamma \rightarrow \mathrm{hadrons}$ (excluding $t\bar{t}$).}
\label{fig:major}
\end{figure}

\begin{figure}
\centering
  \includegraphics[width=0.6\textwidth]{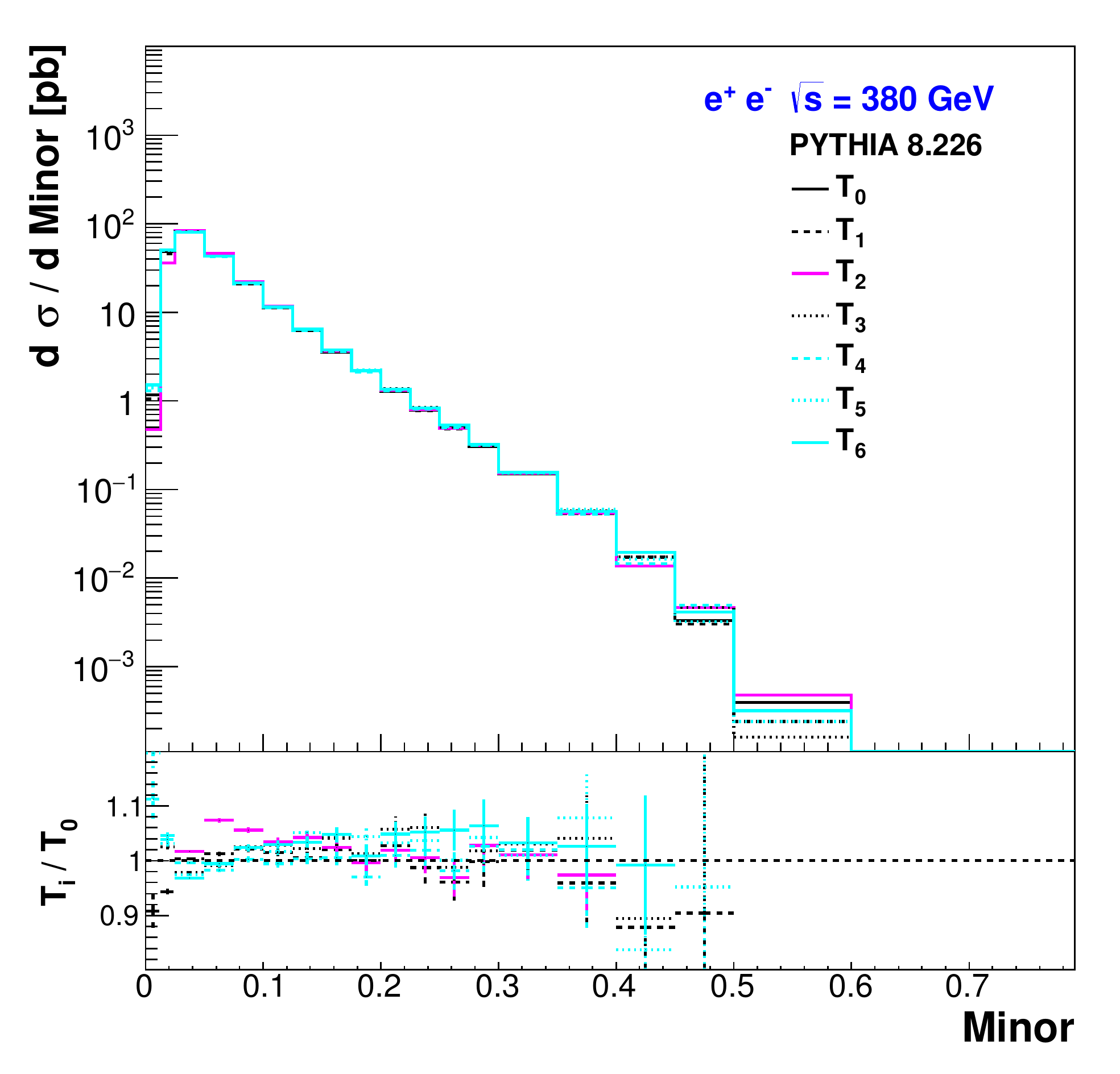}
  \includegraphics[width=0.6\textwidth]{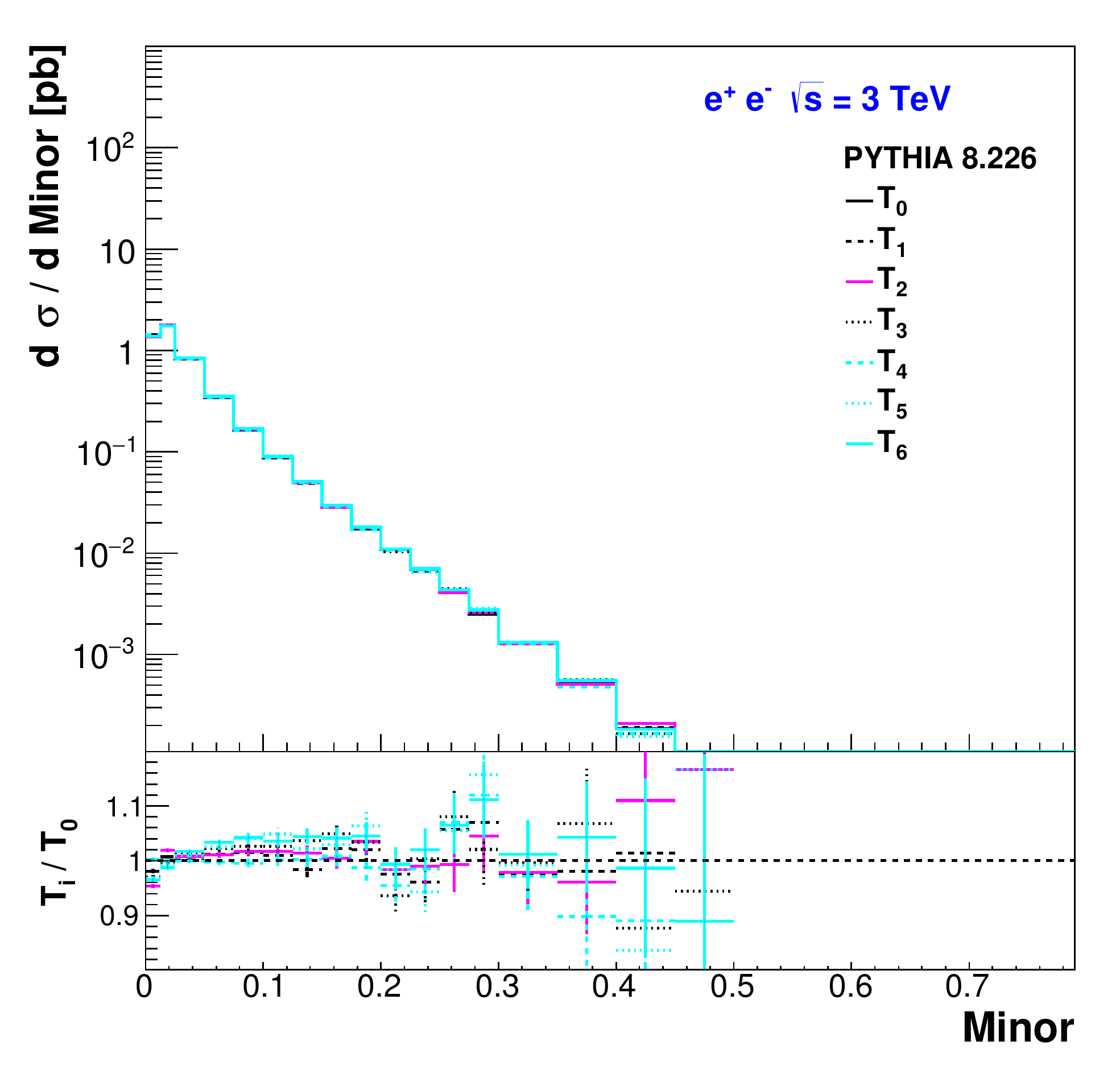}
\caption{The distribution of the minor thrust values for different PYTHIA8 tunes at 380~GeV and 3~TeV CM energies  in   $e^+e^-\rightarrow Z^*/\gamma \rightarrow \mathrm{hadrons}$ (excluding $t\bar{t}$).}
\label{fig:minor}
\end{figure}

\begin{figure}
\centering
  \includegraphics[width=0.6\textwidth]{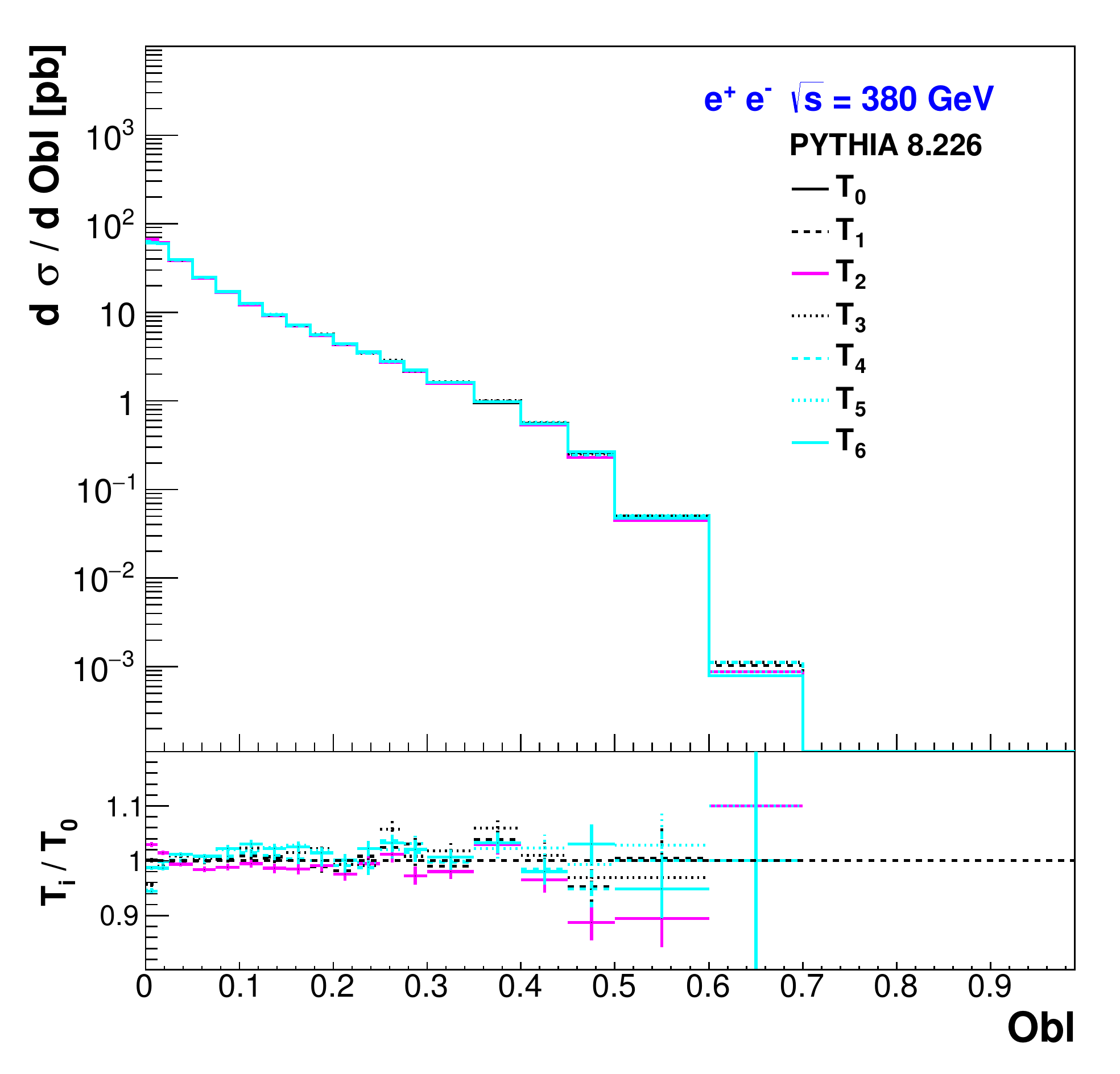}
  \includegraphics[width=0.6\textwidth]{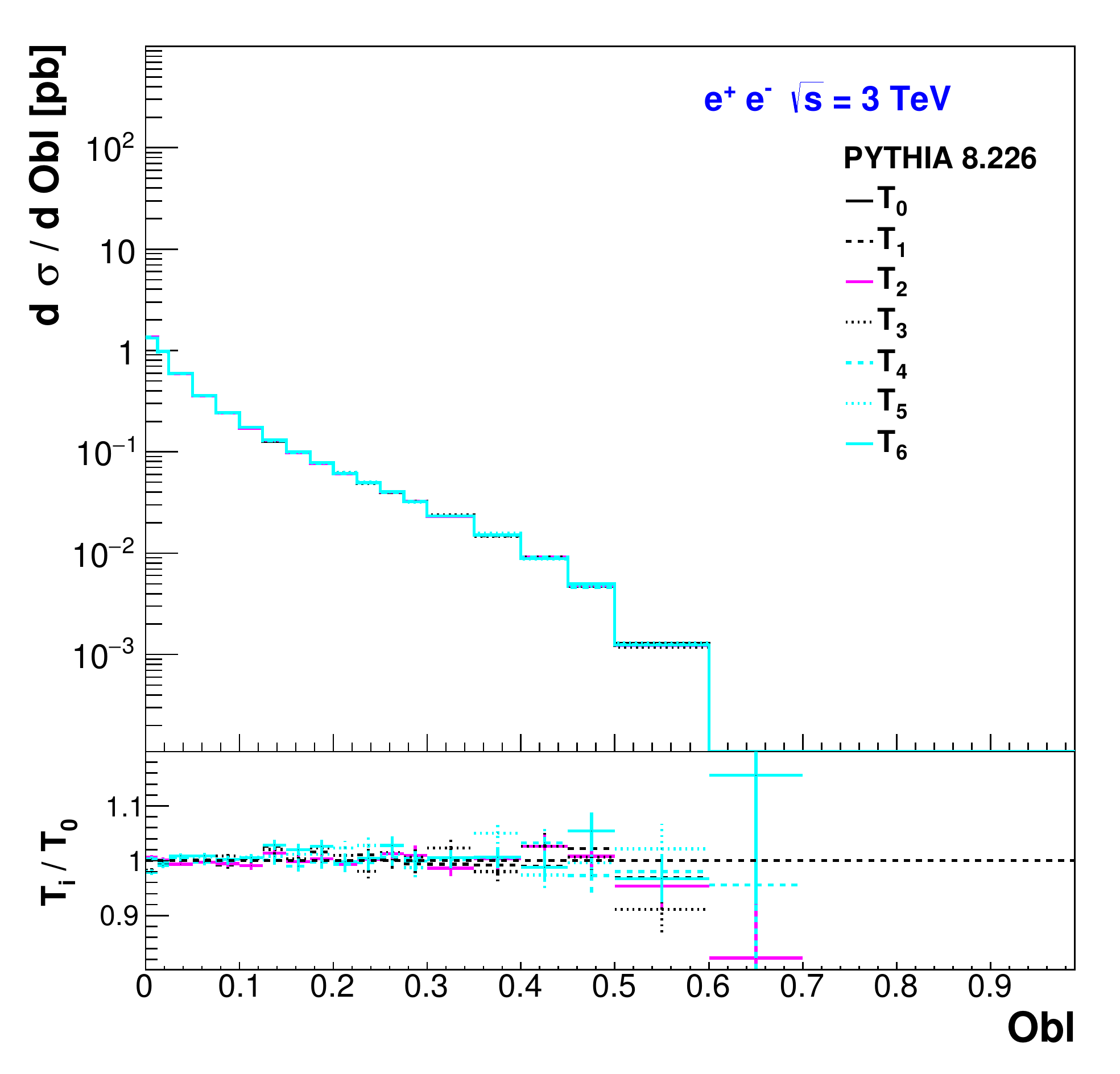}
\caption{The distribution of the oblateness for different PYTHIA8 tunes at 380~GeV and 3~TeV CM energies  in   $e^+e^-\rightarrow Z^*/\gamma \rightarrow \mathrm{hadrons}$ (excluding $t\bar{t}$).}
\label{fig:oblateness}
\end{figure}

%% event shapes in top production
\begin{figure}
\centering
  \includegraphics[width=0.6\textwidth]{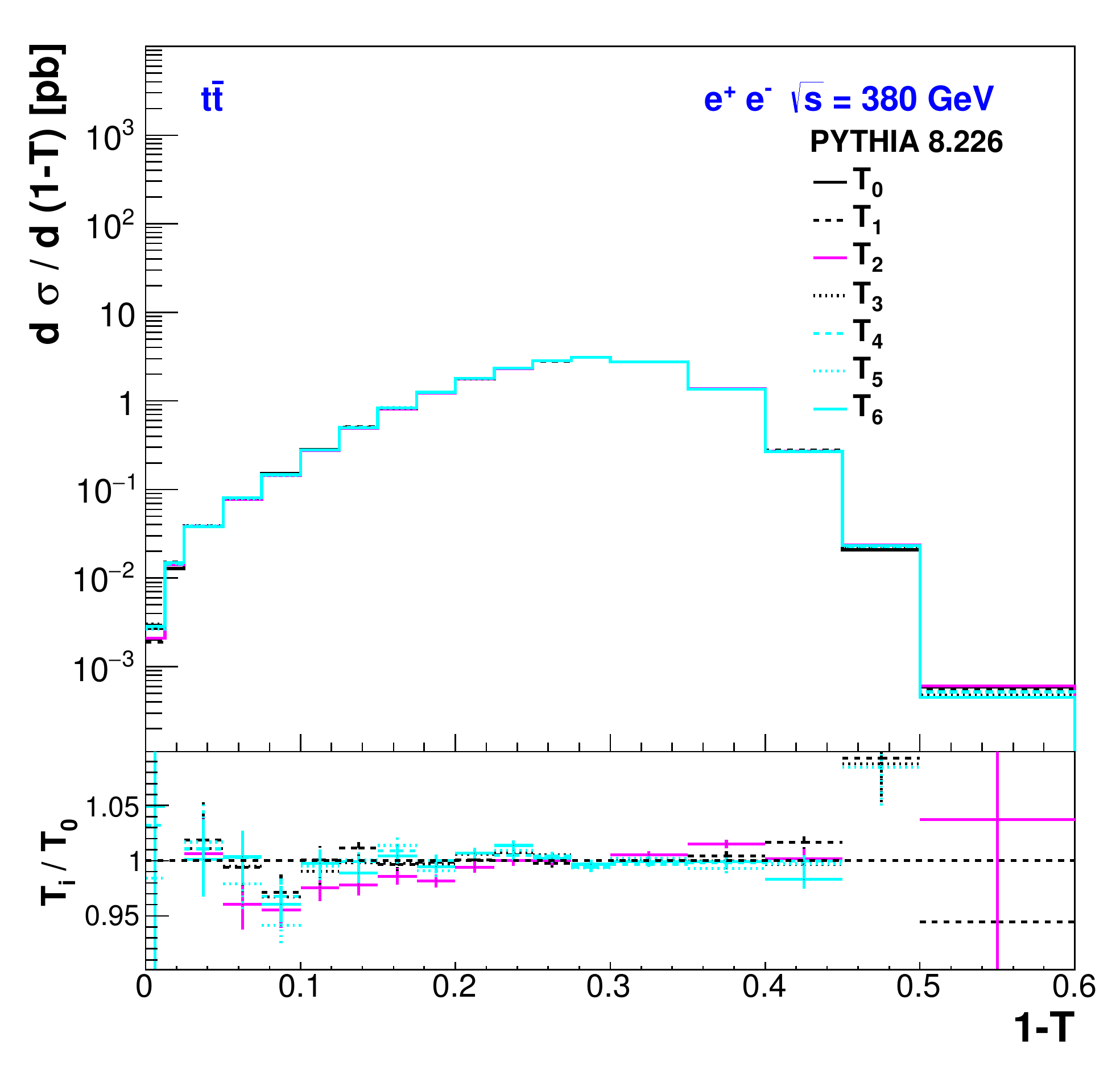}
  \includegraphics[width=0.6\textwidth]{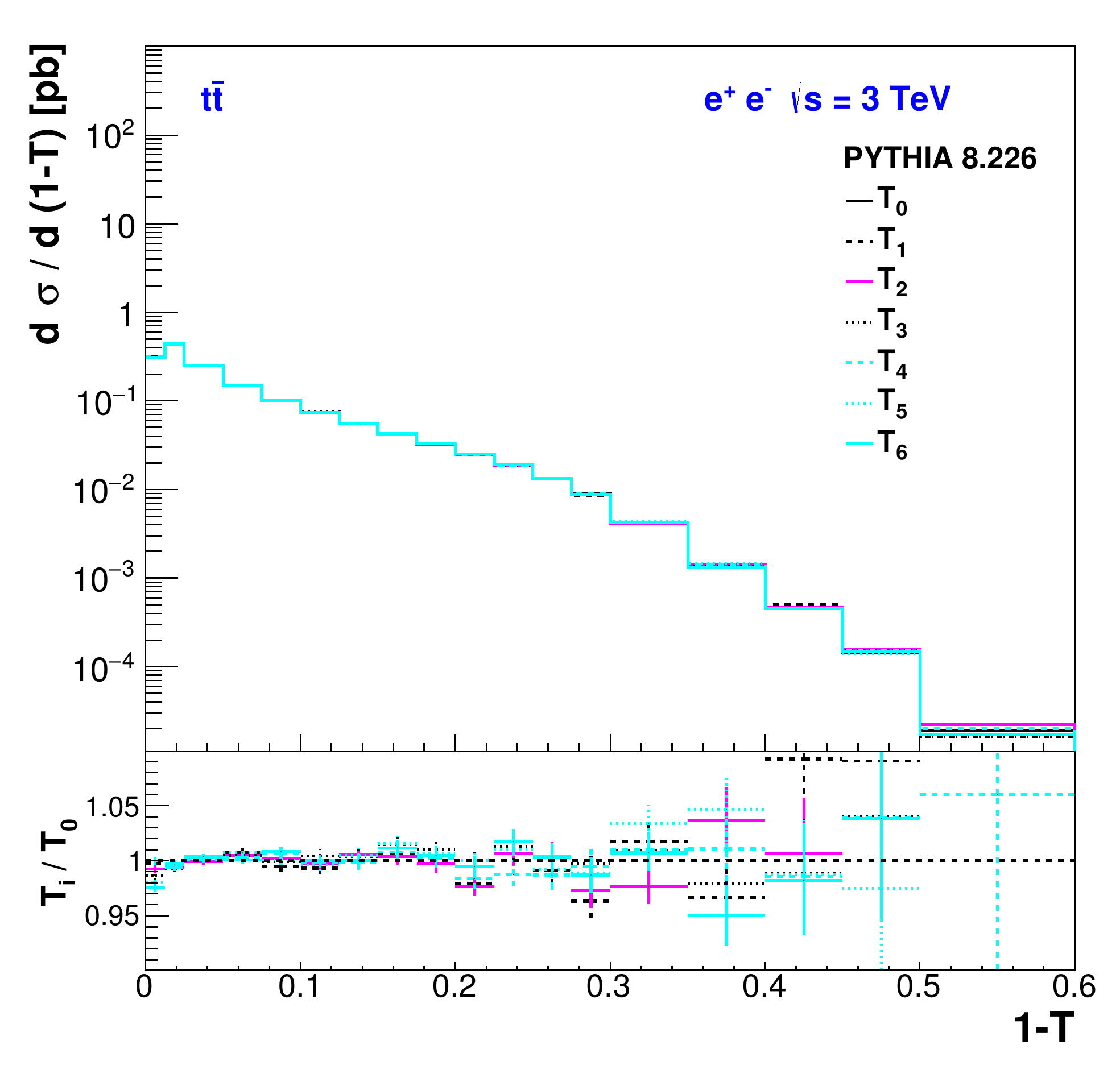}
\caption{The distribution of the thrust values ($1-T$) for different PYTHIA8 tunes at 380~GeV and 3~TeV CM energies in  $e^+e^-\rightarrow Z^*/\gamma \rightarrow t\bar{t}\mathrm{(all\> decays)}$. }
\label{fig:thrust_ttbar}
\end{figure}

\begin{figure}
\centering
  \includegraphics[width=0.6\textwidth]{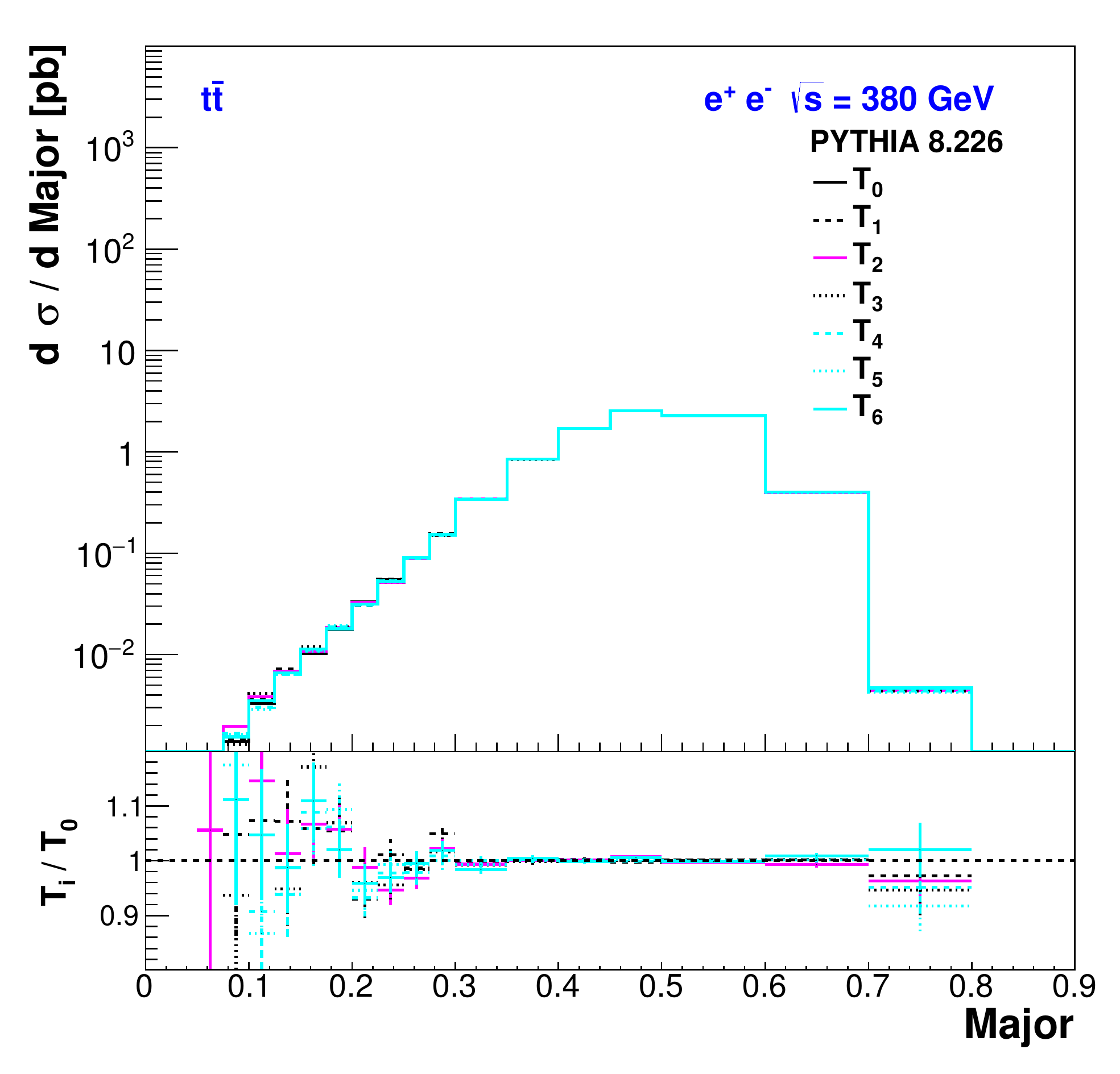}
  \includegraphics[width=0.6\textwidth]{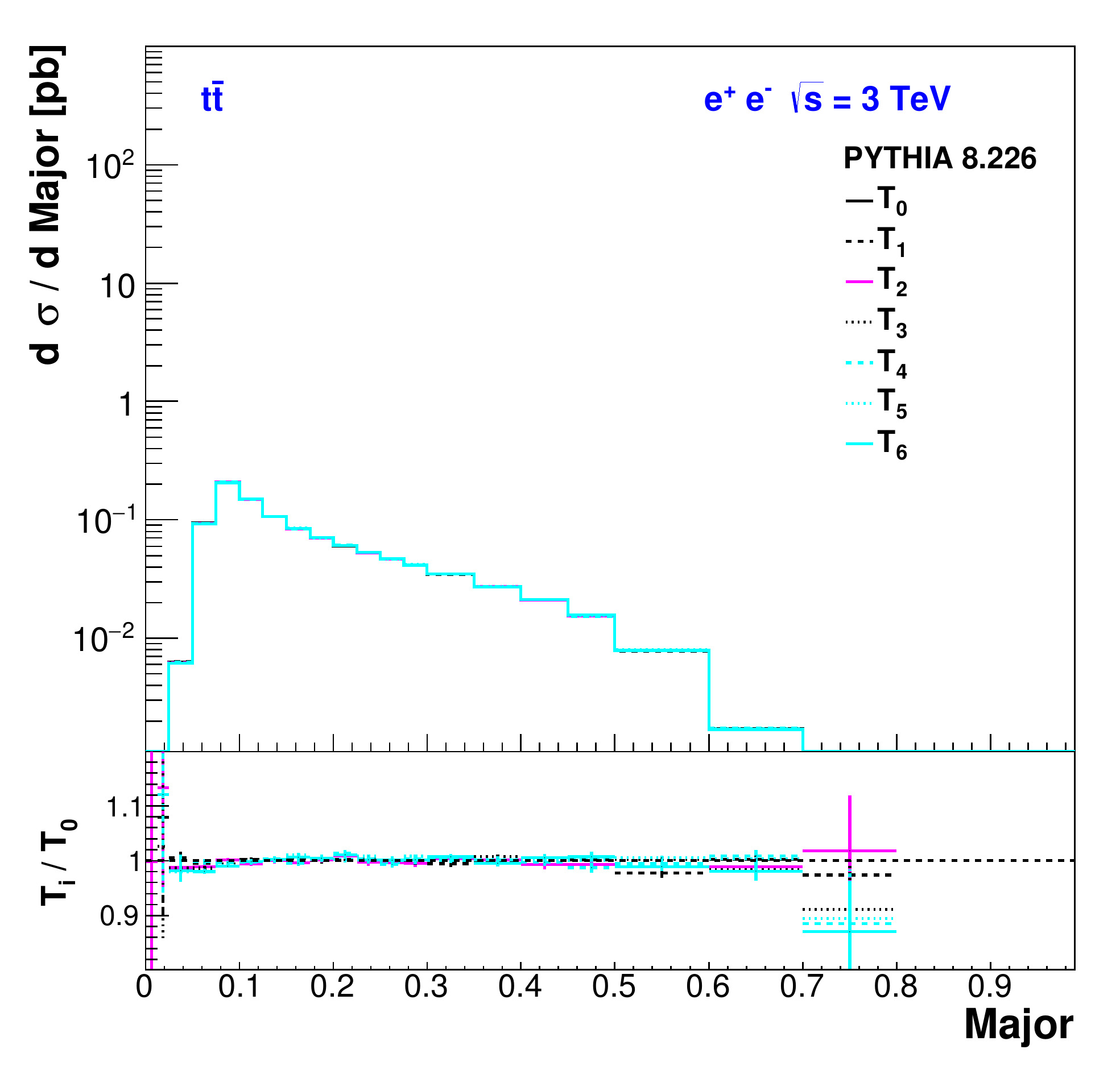}
\caption{The distribution of the major thrust values for different PYTHIA8 tunes at 380~GeV and 3~TeV CM energies in  $e^+e^-\rightarrow Z^*/\gamma \rightarrow t\bar{t}\mathrm{\>(all\> decays)}$.}
\label{fig:major_ttbar}
\end{figure}

\begin{figure}
\centering
  \includegraphics[width=0.6\textwidth]{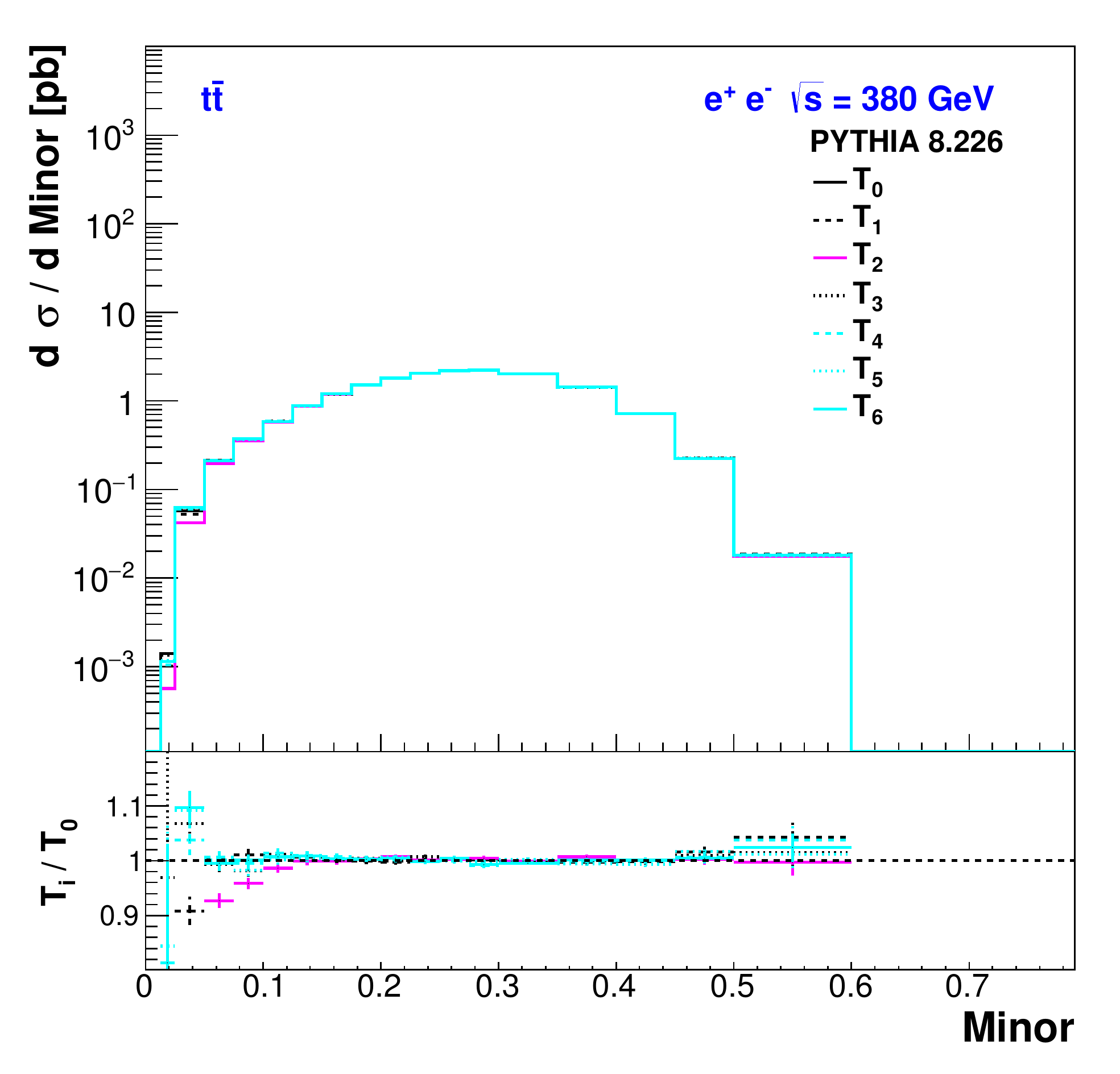}
  \includegraphics[width=0.6\textwidth]{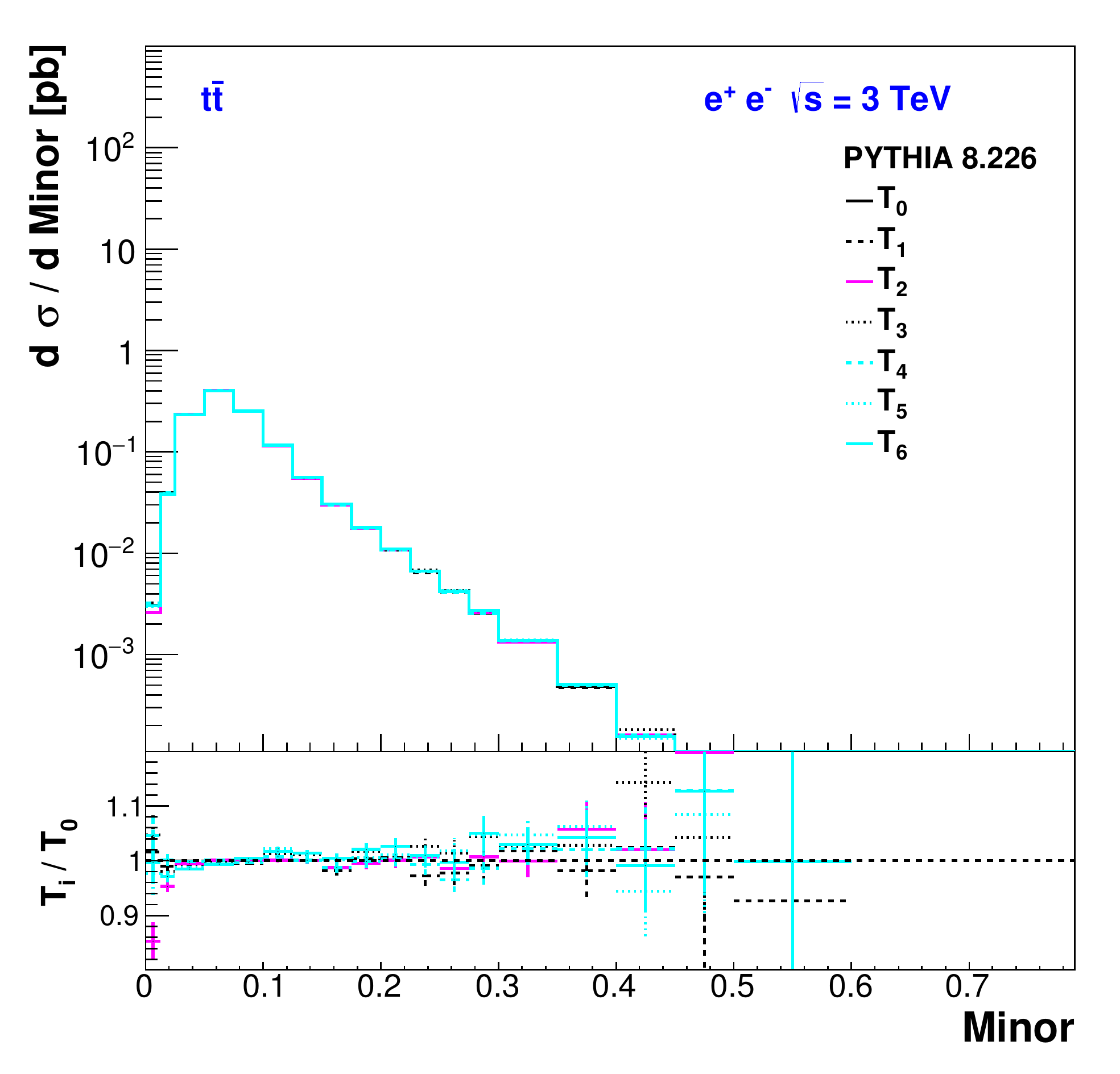}
\caption{The distribution of the minor thrust values for different PYTHIA8 tunes at 380~GeV and 3~TeV CM energies in $e^+e^-\rightarrow Z^*/\gamma \rightarrow t\bar{t}\mathrm{\>(all\> decays)}$.}
\label{fig:minor_ttbar}
\end{figure}

\begin{figure}
\centering
  \includegraphics[width=0.6\textwidth]{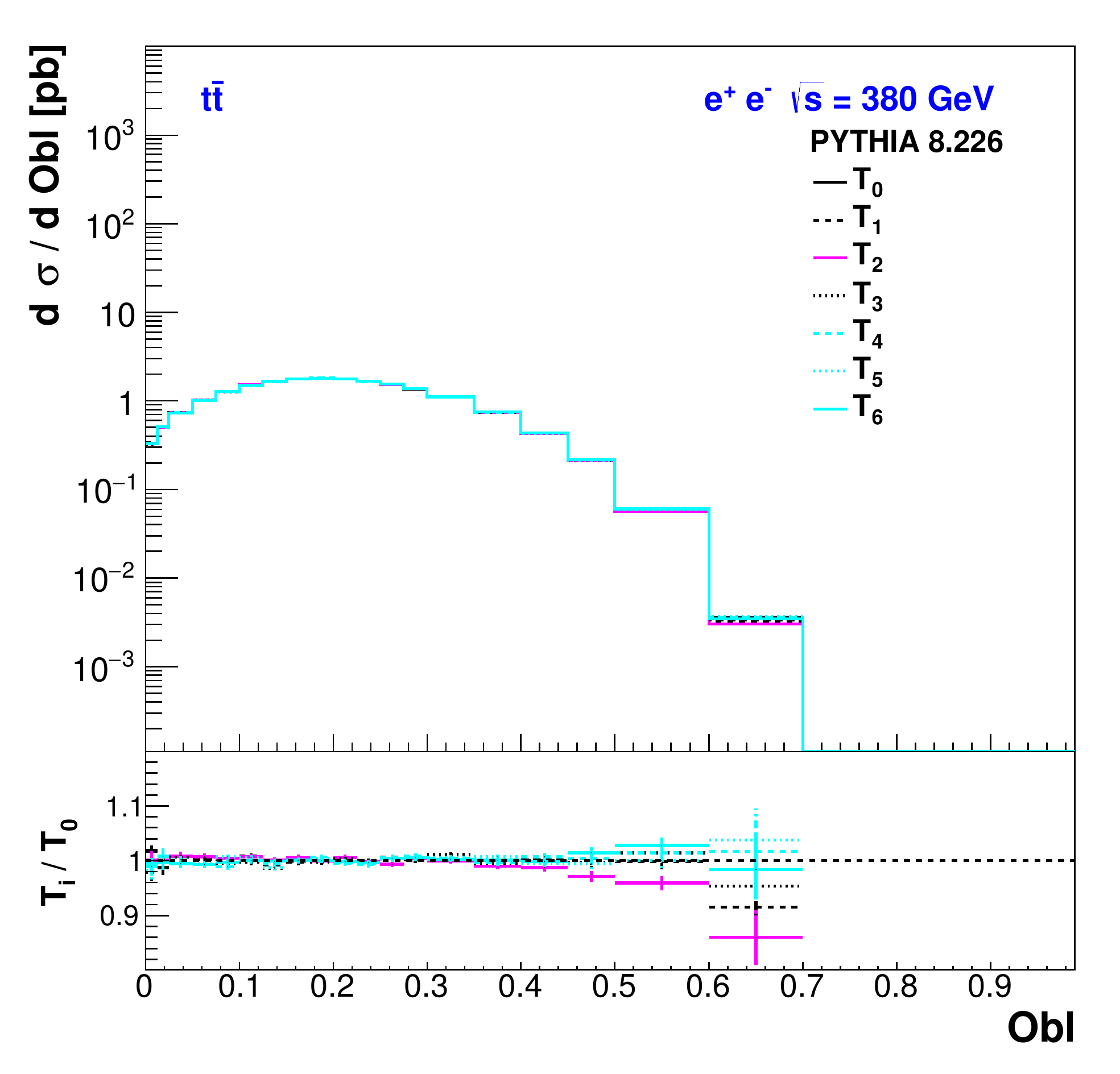}
  \includegraphics[width=0.6\textwidth]{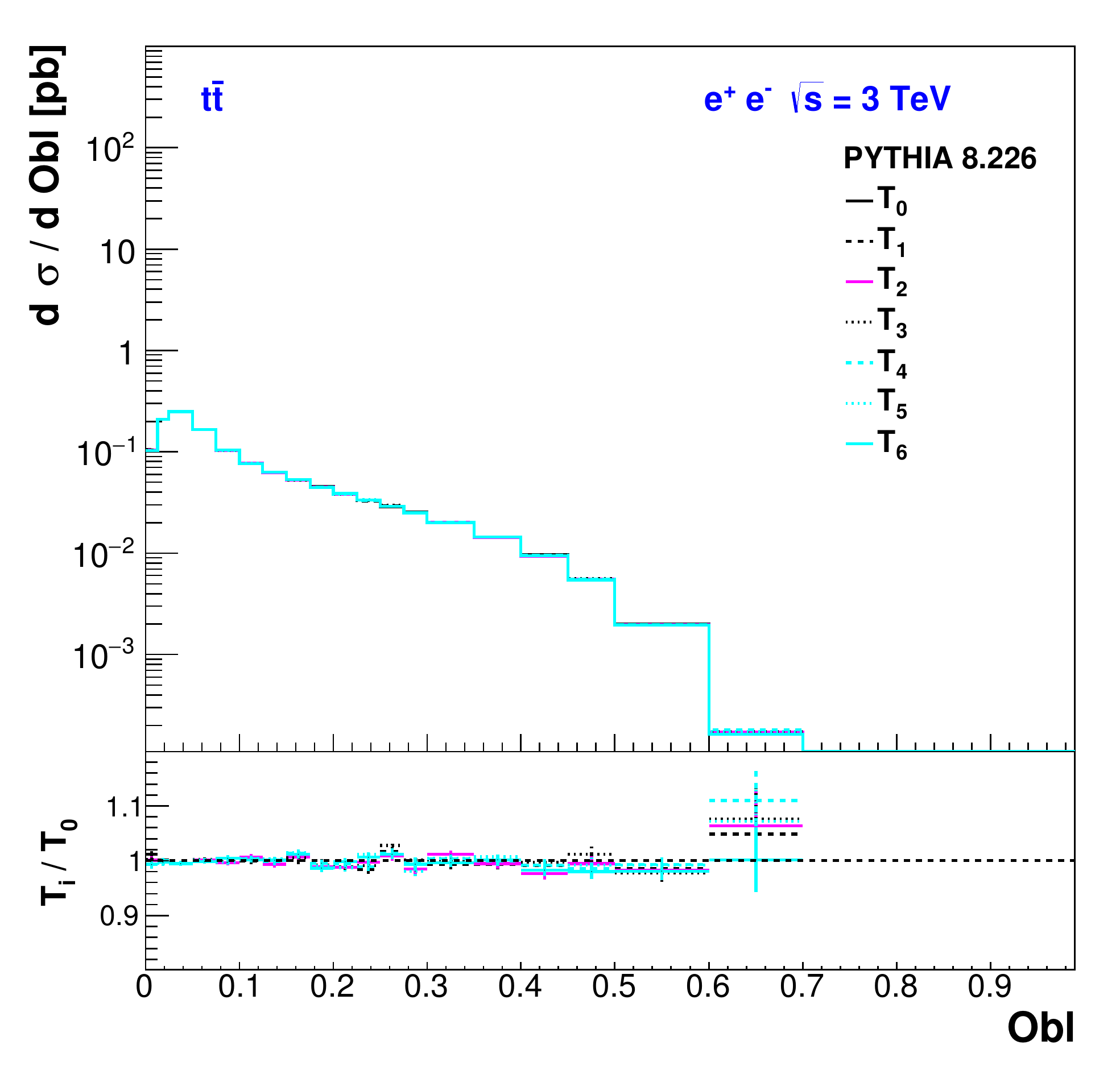}
\caption{The distribution of the oblateness for different PYTHIA8 tunes at 380~GeV and 3~TeV CM energies  in  $e^+e^-\rightarrow Z^*/\gamma \rightarrow t\bar{t}\mathrm{\>(all\> decays)}$.}
\label{fig:oblateness_ttbar}
\end{figure}

%%%%%%%%% masses

\begin{figure}
\centering
  \includegraphics[width=0.6\textwidth]{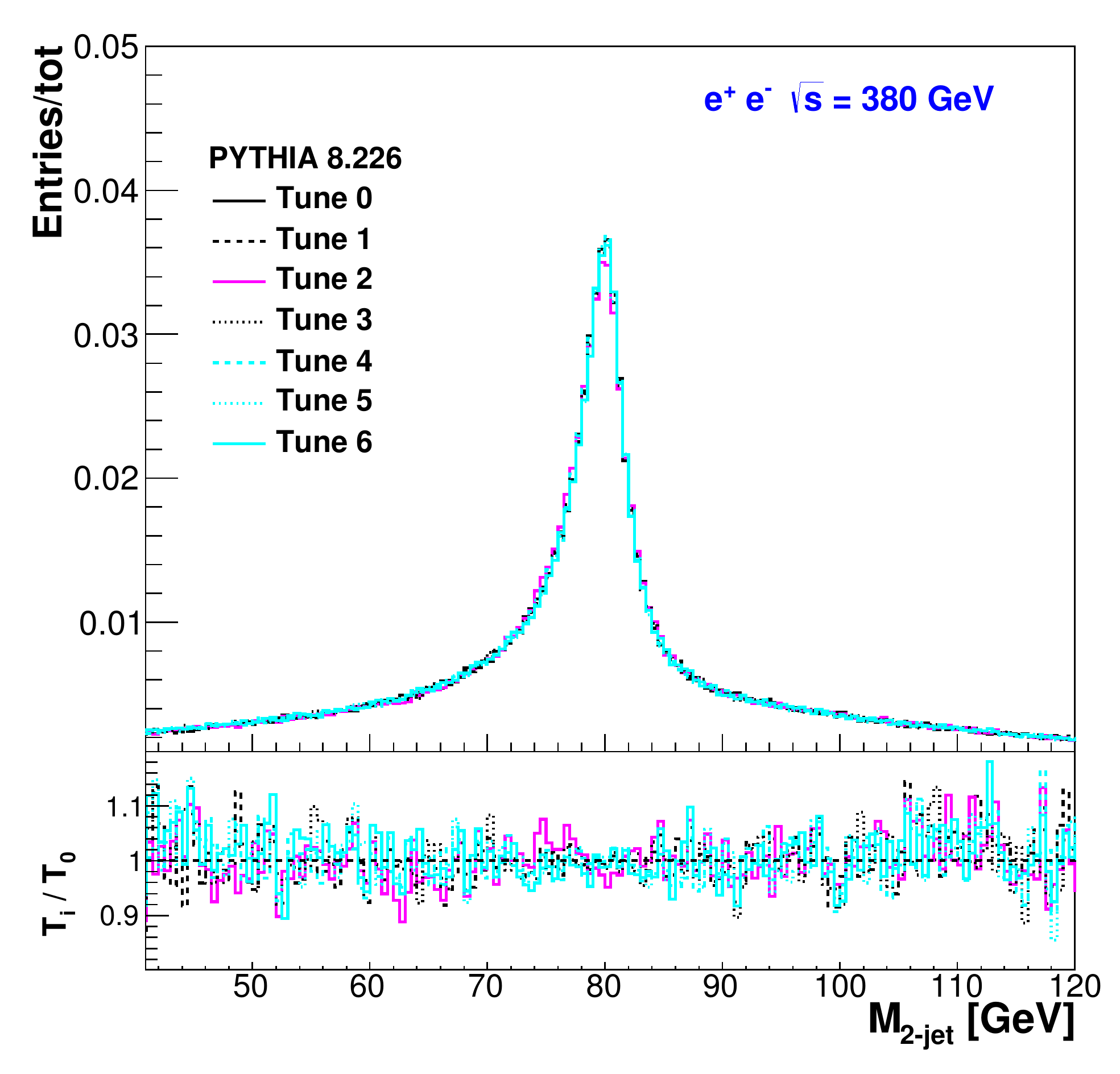}
  \includegraphics[width=0.6\textwidth]{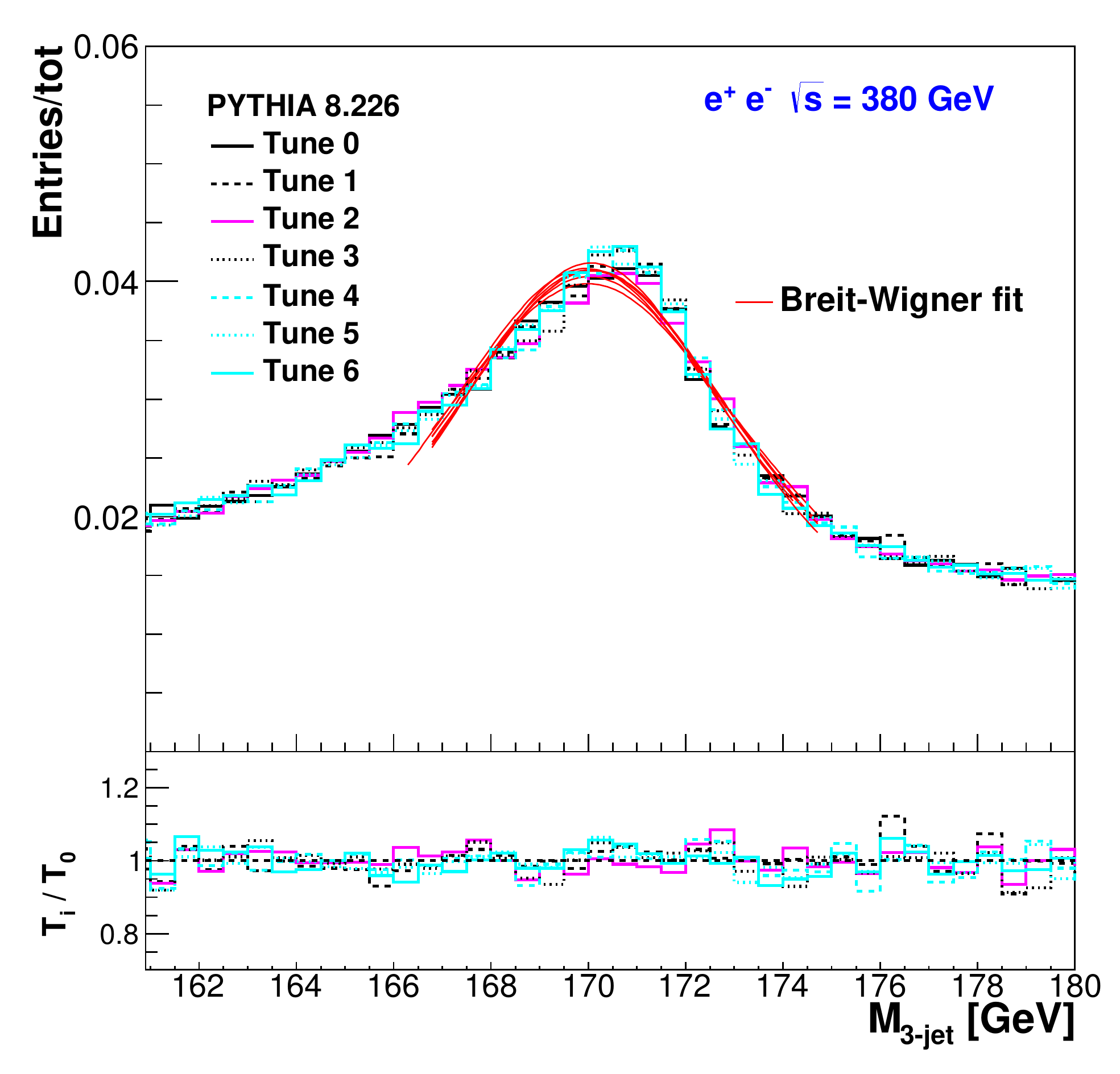}
\caption{Jet invariant masses for two-jet and three-jet combinations at 380~GeV CM energy for semi-leptonic top decays in  $e^+e^-\rightarrow Z^*/\gamma \rightarrow t\bar{t}$.}
\label{fig:resolv}
\end{figure}

\begin{figure}
\centering
  \includegraphics[width=0.65\textwidth]{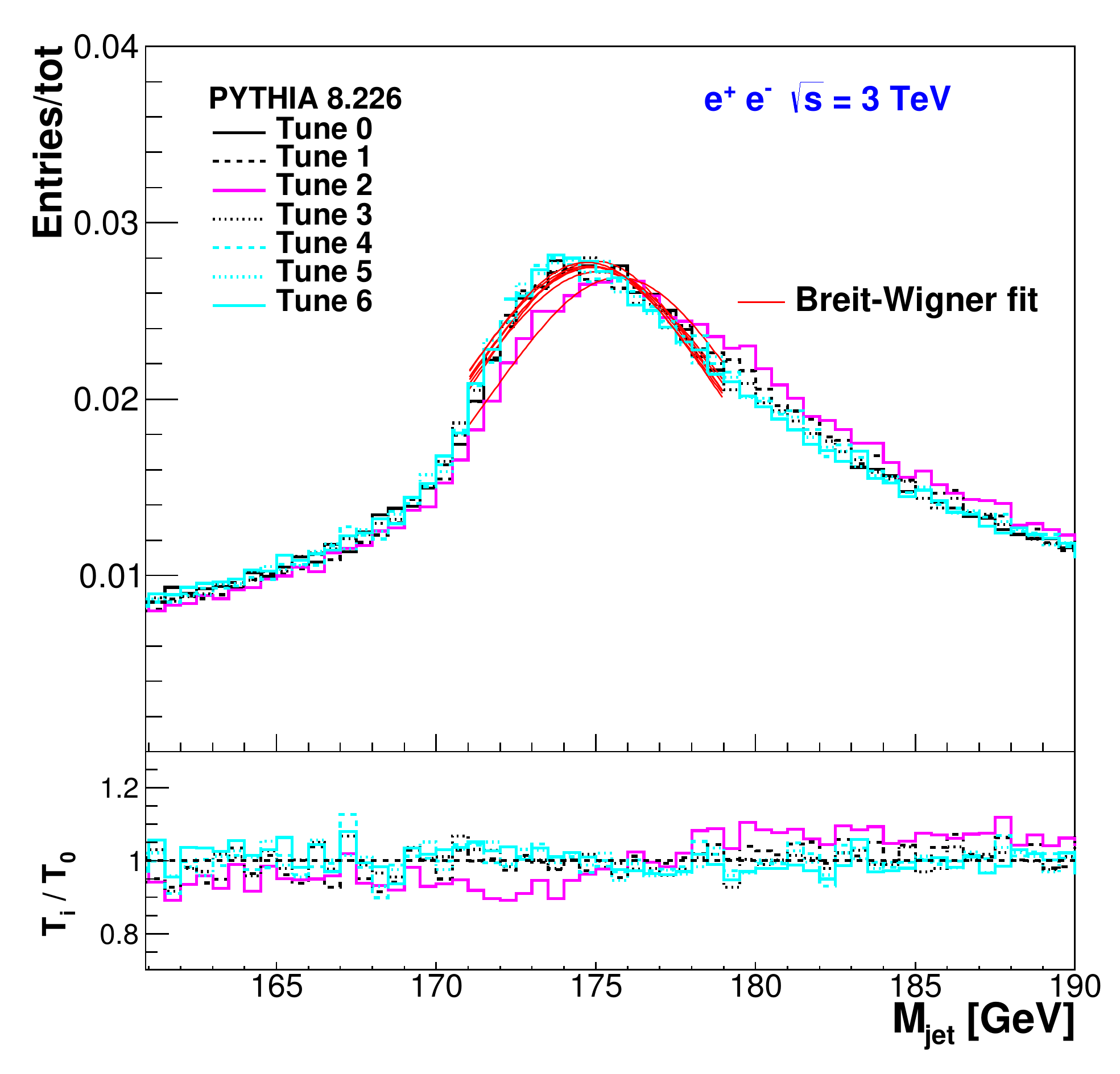}
\caption{Jet masses for boosted top reconstruction in  semi-leptonic top decays at 3~TeV CM energy  in  $e^+e^-\rightarrow Z^*/\gamma \rightarrow t\bar{t}$, together with the fit to determine the peak positions.}
\label{fig:threejet}
\end{figure}